\documentclass[aps,10pt,twocolumn,showpacs,pra,citeautoscript,amsmath,amssymb,floatfix,nofootinbib,superscriptaddress,longbibliography]{revtex4-1}

\usepackage{amsmath}
\usepackage{dsfont}
\usepackage{amssymb}
\usepackage{bbm}
\usepackage{amsthm, thm-restate}
\usepackage{physics}
\usepackage[normalem]{ulem}
\usepackage{makecell}
\usepackage{bm}
\usepackage{bbold}
\usepackage{float}

\usepackage{graphicx}

\usepackage{xcolor, soul}

\usepackage{natbib}
\usepackage[linktocpage]{hyperref} 
\usepackage[capitalise]{cleveref}
\crefname{figure}{Figure}{figures}

\usepackage{algpseudocode}
\usepackage{enumerate}

\newcounter{lemmaN}

\newtheorem{lemma}[lemmaN]{Lemma}

\newcounter{lemmaA}

\newcounter{colN}

\newtheorem*{lemma*}{Lemma}
\newtheorem*{theorem*}{Theorem}

\newcommand{\id}{\mathbbm{1}}
\newcommand{\E}{\mathfrak{E}}

\makeatletter
\newcommand*{\balancecolsandclearpage}{%
  \close@column@grid
  \clearpage
  \twocolumngrid
}
\makeatother

\makeatletter
\def\tocdepth@fullmunge{%
\let\l@section@saved\l@section
\let\l@section\@gobble@tw@
\let\l@subsection@saved\l@subsection
\let\l@subsection\@gobble@tw@
}%
\def\tocdepth@fullrestore{%
\let\l@section\l@section@saved
\let\l@subsection\l@subsection@saved
}%
\newcommand{\hidetoc}[0]{\addtocontents{toc}{\string\tocdepth@fullmunge}}
\newcommand{\restoretoc}[0]{\addtocontents{toc}{\string\tocdepth@fullrestore}}
\makeatother

\begin{document}

\title{Certified randomness from quantum speed limits}

\author{Caroline L.\ Jones}
\email{CarolineLouise.Jones@oeaw.ac.at}
\thanks{}
\affiliation{Institute for Quantum Optics and Quantum Information,
Austrian Academy of Sciences, Boltzmanngasse 3, A-1090 Vienna, Austria}
\affiliation{Vienna Center for Quantum Science and Technology (VCQ), Faculty of Physics, University of Vienna, Vienna, Austria}
\author{Albert\ Aloy}
\email{Albert.Aloy@oeaw.ac.at}
\thanks{\newline AA and CLJ contributed equally to this work.}
\affiliation{Institute for Quantum Optics and Quantum Information,
Austrian Academy of Sciences, Boltzmanngasse 3, A-1090 Vienna, Austria}
\affiliation{Vienna Center for Quantum Science and Technology (VCQ), Faculty of Physics, University of Vienna, Vienna, Austria}
\author{Gerard Higgins}
\affiliation{Institute for Quantum Optics and Quantum Information,
Austrian Academy of Sciences, Boltzmanngasse 3, A-1090 Vienna, Austria}
\affiliation{Marietta Blau Institute for Particle Physics,
Austrian Academy of Sciences, Dominikanerbastei 16, A-1010 Vienna, Austria}
\author{Markus P.\ M\"uller}
\affiliation{Institute for Quantum Optics and Quantum Information,
Austrian Academy of Sciences, Boltzmanngasse 3, A-1090 Vienna, Austria}
\affiliation{Vienna Center for Quantum Science and Technology (VCQ), Faculty of Physics, University of Vienna, Vienna, Austria}
\affiliation{Perimeter Institute for Theoretical Physics, 31 Caroline Street North, Waterloo, Ontario N2L 2Y5, Canada}

\date{January 28, 2026}

\begin{abstract}
Quantum speed limits are usually regarded as fundamental restrictions, constraining the amount of computation that can be achieved within some given time and energy. Complementary to this intuition, here we show that these limitations are also of operational value: they enable the secure generation of certified randomness. We consider a prepare-and-measure scenario with some (experimentally determined or promised) upper bound on the energy uncertainty $\Delta E$ of the average prepared quantum state, but without any further assumptions on the devices, Hilbert space or Hamiltonian. Given that we can freely choose the time at which to apply the untrusted preparation procedure, we show that this scenario admits the generation of randomness that is secure against adversaries with additional classical information. We show how to determine the amount of certified randomness given the observed correlations, discuss how interactions with the environment are taken into account, and sketch a conceivable experimental implementation. In particular, we show that single-mode coherent states admit this kind of certification of non-zero randomness in some parameter regimes, reinforcing existing demonstrations of nonclassicality in the simple harmonic oscillator. Our results extend earlier efforts to devise semi-device-independent protocols grounded in reasonable physical assumptions, and they contribute to the understanding of time-energy uncertainty relations via their operational consequences.
\end{abstract}

\maketitle

\section{Introduction}

Historically, the probabilistic nature of quantum mechanics has been the root of scientific disquiet -- with Einstein famously remarking that ``[Nature] does not throw dice''~\cite{BornEinstein}. Nevertheless, converse to this intuition, the role of indeterminacy became further entrenched in quantum theory with Heisenberg's posited uncertainty principles~\cite{Heisenberg1,Heisenberg2}, which claimed that pairs of canonical variables could only be known up to some jointly bounded precision:
\begin{equation}
\begin{split}
   \Delta x \Delta p &\geq \hbar/2, \\ 
   \Delta E \Delta t &\geq \hbar/2. 
\end{split}
\end{equation}
Whilst the position-momentum uncertainty relation was soon established mathematically~\cite{Bohr}, its time-energy analogue proved more cumbersome. In particular, without a formulation of time as a quantum observable, it was initially unclear how to interpret $\Delta t$ as ``time uncertainty''. An alternative paradigm was proposed by Mandelstam and Tamm~\cite{Mandelstam}, reformulating the relation as a \textit{quantum speed limit} (QSL):
\begin{equation}\label{QSL}
    \Delta t\geq\tau_{\text{QSL}}:= \frac{\pi\hbar}{2\Delta E},
\end{equation}
\begin{figure}[H]
\centering 
\includegraphics[trim=120 135 230 80,clip, width=1.0\columnwidth]{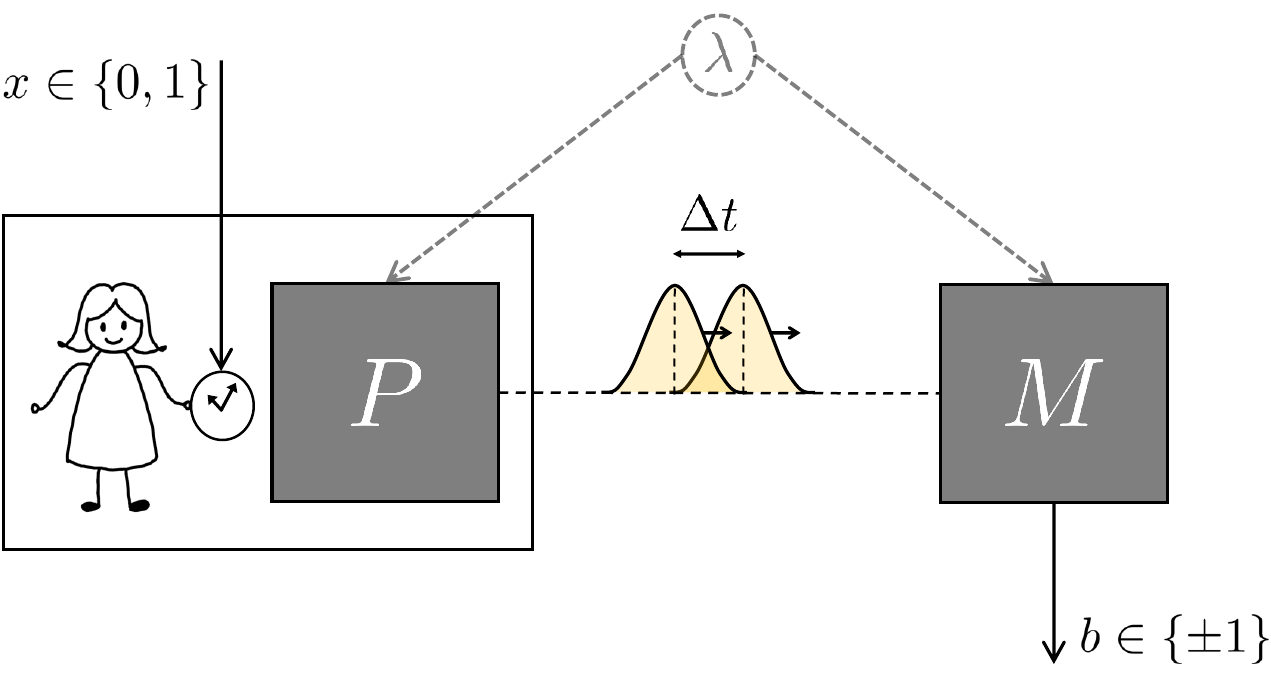}
\caption{The preparation device $P$ prepares a fixed quantum state which is subsequently sent to a measurement device $M$. We treat both $P$ and $M$ as black boxes, but we assume that we have full control of the time $t_x$  at which $P$ is triggered: at $t_0$ or $t_1=t_0+\Delta t$, where $x\in\{0,1\}$ labels the input to $P$. In contrast, $M$ is implemented at a predetermined time. A classical variable $\lambda$ is allowed to influence and possibly correlate $P$ and $M$. It may be known to an adversary, but it is unknown to the user of the randomness generator.}
\label{fig:setup}
\end{figure}
\noindent
specifying a minimum time $\tau_{\text{QSL}}$ that a system takes to evolve to an orthogonal state. The Mandelstam-Tamm bound was further generalised to arbitrary states~\cite{Anandan}:
\begin{equation}\label{QSLpure}
    \Delta t\geq\frac{\hbar\arccos{|\braket{\psi_1}{\psi_0}|}}{\Delta E},
\end{equation}
thereby bounding how fast a system can evolve between any two states $\ket{\psi_0}$ and $\ket{\psi_1}$, where $\Delta E=\Delta E_{\psi_0}=\Delta E_{\psi_1}$ is the standard deviation of the energy in the initial and final states. Note that when the states are fully distinguishable (i.e.\ $\arccos{|\braket{\psi_1}{\psi_0}|}=\pi/2$), Eq.~(\ref{QSL}) is recovered from Eq.~(\ref{QSLpure}). Complementary to Mandelstam and Tamm's bound, Margolus and Levitin~\cite{Margolus,Levitin} derived an alternative formulation of $\tau_{\text{QSL}}$ in terms of the expectation value of the Hamiltonian: $\tau_{\text{QSL}}=\pi\hbar/(2\langle E\rangle)$, where the ground state energy is set to zero.
For a more comprehensive review on QSLs, we refer the reader to~\cite{Deffner-Campbell}.

Whilst typically viewed as a \textit{limitation}~\cite{Bekenstein,Lloyd,Deffner-Clausius,Caneva,Giovannetti1}, in this paper we demonstrate the utility of the QSL for quantum information protocols. In particular, we show its application for randomness generation. At first sight, it seems almost trivial to go from uncertainty relations to randomness: why not just prepare a particle in some state with very small $\Delta p$, ensuring large $\Delta x$, and measure its position? Or why do we not simply send a photon on a half-silvered mirror and record the outcome? A reason to be careful is that we may not trust our devices (the mirror, or the state preparation procedure): perhaps these device are producing outcomes that look random to us, but are in fact predictable by an adversary who holds additional information about the state of the system.

These restrictions can be overcome with device-independent or \textit{semi-device-independent} protocols. In the latter case, we have a prepare-and-measure scenario (such as the one in Figure~\ref{fig:setup}) where we do not trust the preparation and measurement devices $P$ and $M$ and treat them as black boxes, but we trust the validity of an assumption about the physical system that is sent from $P$ to $M$. For example, we may assume that the quantum system to be sent is a qubit, or, more generally, that it is described by a Hilbert space of some fixed dimension $d$~\cite{gallego2010device,bowles2014certifying,brunner2008testing,PawlowskiBrunner,Li2011,Li2012,Mironowicz2016}. Observing some correlations between the inputs of $P$ (and perhaps $M$), and the output of $M$, we can then ensure that this output must be at least partially random, even relative to every adversary holding an arbitrary amount of additional classical information about the devices $P$ and $M$ and all variables that are physically relevant for these procedures.

However, the assumption of Hilbert space dimension $d$ is arguably not very well physically motivated. Therefore, various protocols have been proposed which replace the dimensionality assumption by assumptions that have a more direct physical or information-theoretic meaning. Several works have analysed semi-device-independent schemes based on assumptions about the overlaps between the prepared states~\cite{BohrBrask2017,Tebyanian2021,RochiCarceller2024,Ioannou2022}, or about the information content of the transmitted system~\cite{tavakoli2020informationally,TZC2022}, the amount of contextuality~\cite{Flatt,Carceller}, or fidelity with the target states~\cite{Tavakoli2021} with important conceptual and formal relations between the different proposals~\cite{Pauwels}. In particular, van Himbeeck et al.~\cite{VanHimbeeck} have shown that assuming a bound on the expected energy $\langle E\rangle$ of the prepared state admits the certification of randomness~\cite{VanHimbeeck2}. However, this assumes that the corresponding Hamiltonian $\hat{H}$ has a unique ground state, and assumes knowledge of the relation between the energy expectation value and the gap above the ground state. Here, we show how randomness can be certified by assuming an upper bound on the energy uncertainty $\Delta E$ of the prepared state, without knowing or assuming anything about the Hamiltonian. Whilst our work is formulated in terms of the energy variance, it may be noted that the Margolus-Levitin bound could also be implemented instead of the Mandelstam-Tamm bound. The former assumes a bound on the energy expectation (as in~\cite{VanHimbeeck,VanHimbeeck2}) rather than the variance, but has the disadvantage that its generalisation to non-orthogonal pairs of pure states does not admit a similarly simple closed-form analytic expression as that of the latter~\cite{Giovannetti2}. This is related to the fact that the energy uncertainty $\Delta E$ is the pure-state special case of the quantum Fisher information, which directly determines the speed of quantum evolution~\cite{Toth}. We therefore restrict our attention to the bound given by Eq.~(\ref{QSLpure}).

While we show that our scheme can generate secure random numbers, it is not primarily the potential technological applicability of this which motivates our work. Instead, we see it as an instance of a more foundational question: how does the structure of space and time constrain the (quantum or more general) correlations that we observe? In other words: if we ``put the black boxes (of DI or semi-DI quantum information) into space and time'', does this give us additional constraints that we may harness or that may potentially explain some of the structure of quantum theory? Here, we find an instance of the former: that choosing the input for the preparation by choosing the time to trigger it allows us to certify a form of nonclassicality without trusting the devices, and with one single further assumption: an upper bound on $\Delta E$.

Conceptually, our scheme builds on the idea that we can perform \textit{trusted operations on untrusted devices}: Even if our preparation procedure is treated as a black box, we typically assume that we have an evident, pretheoretic, ``macroscopic'' notion of some operations that we can apply to it, regardless of the ``microscopic'' details of the device. For example, we think that we know what it means to supply an input to a black box, or to rewire the outputs of several boxes, or to place several boxes far apart in spacelike separation (as in Bell experiments). In previous work~\cite{RotationBoxes1}, we have assumed that we know how to rotate preparation devices around a fixed axis in space, and we have analyzed the resulting prepare-and-measure correlations within and beyond quantum physics~\cite{RotationBoxes2}.
Similarly, here we imagine that we can freely decide the time at which we operate a given box (say, by pressing the button that triggers the untrusted preparation procedure), which yields a semi-device-independent randomness generation scheme based solely on an upper bound to the system's energy standard deviation $\Delta E$.

We emphasise that our results build to a significant extent on earlier works by van Himbeeck et al.~\cite{VanHimbeeck,VanHimbeeck2}, but with some important differences and novelties. For example, the fact that the standard deviation $\Delta E$ is not linear implies that the ``classical max average'' set of correlations $\mathcal{\bar C}_{\E,\Delta t}$ is not a subset of the quantum set $\mathcal{Q}_{\E,\Delta t}$, and that their algorithm must be adapted to determine the amount of certified randomness. The use of quantum speed limits leads to the necessity of further physical and conceptual considerations. See Section~\ref{SecOutlook} for a more detailed comparison to existing work.\\

\textbf{Our article is organised as follows.} In Section~\ref{SecDescription}, we give a theoretical description of the prepare-and-measure scenario. We characterise the set of correlations that are consistent with quantum theory for any given $\Delta E$ and $\Delta t$, and use concavity of the variance to determine the set of correlations that are ``classical'', i.e.\ that do not admit the generation of certified randomness. We show how the results generalise from closed system evolution to interactions with an environment. In Section~\ref{randomness}, we show how the amount of certifiable randomness can be determined, and give an example plot of the result. In Section~\ref{SecExperimental}, we describe a possible experimental implementation involving the quantum harmonic oscillator, and we conclude in Section~\ref{SecOutlook}.

\section{Theoretical description}
\label{SecDescription}

We follow the semi-device-independent (semi-DI) protocol of~\cite{VanHimbeeck}, consisting of a simple prepare-and-measure scenario as depicted in Figure~\ref{fig:setup}. We begin by describing the scenario without taking the variable $\lambda$ into account. A preparation box $P$ takes an input $x\in\{0,1\}$, and sends some system to a measurement box $M$, which yields one of two outputs $b\in\{\pm 1\}$. By construction, the only effect of the input is to control the time at which the preparation procedure is implemented: in the case of input $x=0$, the state is prepared at time $t_0$, or, in the case $x=1$, the state is prepared at time $t_1=t_0+\Delta t$. While we do not trust the device $P$, we assume that we can fully control the time at which we perform $P$, similarly as we believe that we can control the choice of input or whether we implement any operation at all. Thus, the two possible states $\rho_x$ that may be sent to $M$ are time-displaced with respect to one another by some delay $\Delta t$.

The measurement device then produces outcome $b$, and is described by some POVM $\{M_b\}$. Minimal assumptions are made about the devices, therefore $\rho_x$ and $M_b$ are treated as unknown and may fluctuate according to some classical random variable(s) $\lambda$, unbeknownst to the experimenter. This variable $\lambda$ can be thought of as containing additional information about the world, which could in principle constitute some predictive advantage for the outcome $b$. Our goal is then to quantify over all possible assignments of $\lambda\in\Lambda$ such that security is guaranteed independently of knowledge of these variables. This means that we allow for shared randomness that is able to correlate the workings of the devices $P$ and $M$. The joint behaviour of the devices is therefore characterised by the probabilities 
\begin{align}
   P(b|x)=\sum_\lambda p(\lambda) \text{tr}\big[M^\lambda_b U_{t_x}\rho^\lambda_0U_{t_x}^\dagger\big],
   \label{EqWithLambda}
\end{align}
with $t_x\in\{0,\Delta t\}$, and $U_{t_x}=e^{-iHt_x/\hbar}$ describes the unitary evolution of the system under some fixed Hamiltonian $\hat{H}$. For now, we assume that the prepared quantum state $\rho_x$ undergoes closed-system evolution according to a fixed Hamiltonian $\hat{H}$ which defines its energy's standard deviation
\[
   \Delta E_{\rho_x}=\sqrt{{\rm tr}(\rho_x \hat{H}^2)-({\rm tr}(\rho_x \hat{H}))^2},
\]
which gives identical values for $x=0$ and $x=1$.
We will relax this assumption further below.

The presence of useful correlations is expressed by deviation from the line $C_0=C_1$, where $C_x$ characterises the bias of the output towards $\pm1$ for the input $x$:
\begin{equation}
    C_x=P(+1|x)-P(-1|x) \quad (x\in\{0,1\}).
\end{equation}
In particular, $C_0\neq C_1$ indicates that the outcome $b$ is influenced by the choice of input $x$. We will now analyse the possible correlations given some value of $\Delta t$ and upper bound on $\Delta E$.

\subsubsection*{Quantum correlations}

Let us begin by considering the prepare-and-measure scenario without the shared randomness $\lambda$, and discuss how to reintroduce it at the end of this section. In this case, we have probabilities $P(b|x)={\rm tr}(M_b \rho_x)$ with $\rho_x=U_{t_x} \rho_x U_{t_x}^\dagger$, such that $C_x={\rm tr}(\rho_x M)$ with $M=M_{+1}-M_{-1}$. Let us initially assume that the preparation device prepares a pure state $|\psi_0\rangle$ at time $t_0$. Then, at time $t_1=t_0+\Delta t$, the state is given by 
\begin{equation}
    \ket{\psi_1}=U_{\Delta t}\ket{\psi_0}=\sum_ne^{-iE_n\Delta t/\hbar}c_n\ket{E_n},
\end{equation}
where $\ket{\psi_0}=\sum_nc_n\ket{E_n}$ is decomposed into its energy eigenbasis (we absorb all further time evolution into the definition of the measurement procedure $M$).

We define the quantum set of pure state correlations for time delay $\Delta t\geq 0$ and maximal energy uncertainty $\E\geq 0$ as follows:
\begin{align}
\mathcal{Q}_{\E, \Delta t}:=\Big\{&(C_0,C_1)\, \big| \, C_x=\bra{\psi_x}M\ket{\psi_x}, -\mathbb{1}\leq M\leq\mathbb{1}, \nonumber\\ 
& \exists \hat{H} \,\text{s.\ t.\ }\, \ket{\psi_1}\!=\!U_{\Delta t}\ket{\psi_0}, \Delta E_{\psi_0}\leq \E\Big\}.
\end{align}
The penultimate condition imposes the existence of some fixed Hamiltonian $\hat{H}$ that relates states $\ket{\psi_0}$ to $\ket{\psi_1}$ via unitary time evolution, where the energy uncertainty of the initial state $\Delta E_{\psi_0}$ is less than some maximal fixed value $\E$. This condition $\Delta E_{\psi_0}\leq \E$ implements the semi-DI assumption for our scenario, expressing our belief of an upper bound on the energy uncertainty. (Here we begin by stating the assumption for pure states, but we will generalise it to mixed states and formulate the exact assumption in more detail below.)

\begin{figure}[t]
\centering 
\includegraphics[trim=0 0 400 0,clip,width=0.9\columnwidth]{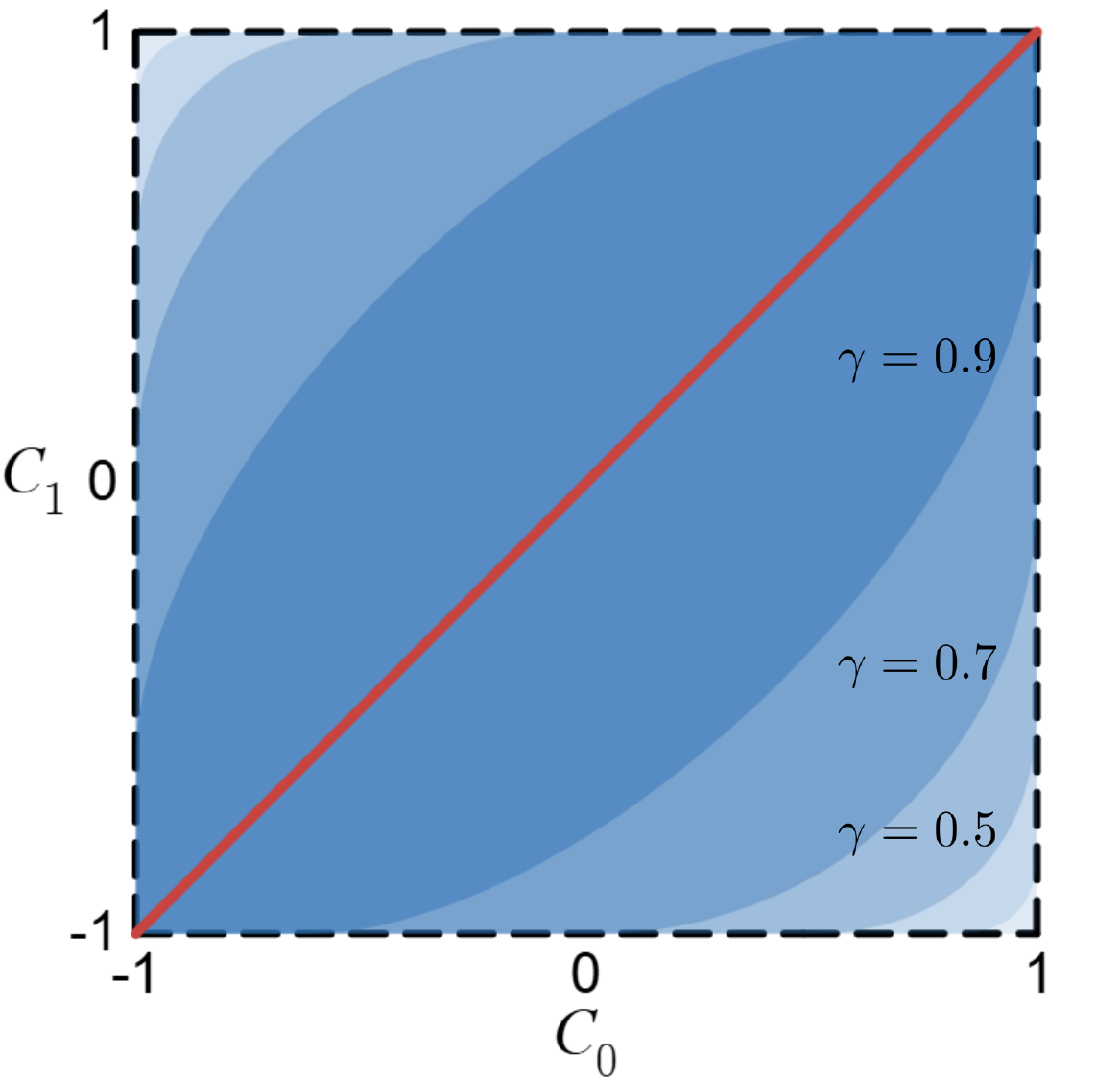}
\caption{Set of quantum (blue) and classical (red) correlations for different state overlaps $\gamma \in \{0.9,0.7,0.5,0.3,0.1\}$  (from darkest to lightest).
}
\label{fig:correlations}
\end{figure}

It has already been proven in~\cite{VanHimbeeck} that, in a prepare-and-measure scenario communicating pure states, the possible correlations are characterised by the inequality
\begin{equation}
\frac{1}{2}\left(\sqrt{1+C_0}\sqrt{1+C_1}+\sqrt{1-C_0}\sqrt{1-C_1}\right) \geq \gamma,
\label{quantum-pure}
\end{equation}
where $\gamma$ is defined as the smallest possible overlap between the two states $\ket{\psi_0}$ and $\ket{\psi_1}$. An overlap $\gamma=1$ corresponds to the maximally restricted set of correlations $C_0=C_1$, whilst as $\gamma\rightarrow1$ the range of accessible correlations increases. For our scenario, the quantum speed limit constrains the overlap as follows:
\begin{equation}
\gamma:=\min|\braket{\psi_1}{\psi_0}|=\begin{cases}\cos\left(\E\Delta t\right) & \text{if } \E\Delta t<\frac{\pi}{2},\\ 0 & \text{otherwise.} \end{cases}
    \label{EqGamma}
\end{equation}
(Note that from here, we use natural units $\hbar=1$ for simplicity.)
This allows us to evaluate the set of pure-state quantum correlations, given the energy uncertainty of the state prepared by $P$ and the time delay between possible preparation times. In Appendix~\ref{quantum-model}, we show that all correlations that satisfy Eq.~(\ref{quantum-pure}) with equality (corresponding to the two curves that bound the blue set (for any fixed $\gamma$) in Figure~\ref{fig:correlations} have a quantum model, i.e.\ are contained in $\mathcal{Q}_{\E,\Delta t}$. Furthermore, by considering purifications with an ancilla system, we prove in Appendix \ref{appendix-mixed} that the set of correlations is unchanged when one allows for mixed states, related according to $\rho_1=U_{\Delta t}\rho_0 U_{\Delta t}^\dagger$
and with energy constraint $\Delta E_{\rho_0}\leq \E$. This shows that $\mathcal{Q}_{\E,\Delta t}$ is convex, and thus, that $\mathcal{Q}_{\E,\Delta t}$ is exactly the set of $(C_0,C_1)$ that satisfy~(\ref{quantum-pure}). It may be noted that the boundary of Eq.~(\ref{quantum-pure}) can be reformulated as an ellipse orientated along the line $C_0=C_1$ -- therefore the set is given by the convex hull of the points on the ellipse and the corner points $(1,1)$ and $(-1,-1)$. For the interested reader, a reformulation is given in~\cite{VanHimbeeck}.

Now we will consider the case that there is shared randomness $\lambda$. In this case, the observed correlation $\mathbf{C}=(C_0,C_1)$ arises from probabilities of the form of Eq.~(\ref{EqWithLambda}), i.e.\  $C_x=\sum_\lambda p(\lambda){\rm tr}[M^\lambda \rho_x^\lambda]$, where $M^\lambda=M^\lambda_{+1}-M^\lambda_{-1}$. Our goal is to show that $\mathbf{C}\in\mathcal{Q}_{\E,\Delta t}$ for $\E= \Delta E_{\rho_0}$. To show that there is a quantum realisation of $\mathbf{C}$ as required to be contained in $\mathcal{Q}_{\mathfrak{E},\Delta t}$, consider a larger Hilbert space (of unbounded but finite dimension) and the block matrices $\bar\rho_0:=\bigoplus_\lambda p(\lambda)\rho_0^\lambda$ and $\bar M:=\bigoplus_\lambda M^\lambda$, then $\bar\rho_0$ is a density operator and $-\mathbb{1}\leq \bar M \leq\mathbb{1}$ because $\bar M=\bar M^\dagger$ and all its eigenvalues are in the interval $[-1,1]$. Furthermore, $\bar H:=\bigoplus \hat{H}$ defines a Hamiltonian that evolves each $\lambda$-subspace independently via $\bar U_t=e^{-i\bar H t}=\bigoplus_\lambda e^{-i \hat{H}t}=\bigoplus_\lambda U_t$. For $\bar \rho_x:=\bar U_{t_x} \bar \rho_0 \bar U_{t_x}^\dagger$, we get ${\rm tr}[\bar M \bar\rho_x]=C_x$, i.e.\ we have reproduced the correlation $\mathbf{C}$ in a scenario with a \emph{fixed} measurement $M$. Furthermore, ${\rm tr}(\bar\rho_0 \bar H^k)={\rm tr}(\rho_0 \hat{H}^k)$ for $k\in\{1,2\}$, and so the construction also preserves the energy's standard deviation $\Delta E_{\rho_0}=\Delta E_{\bar \rho_0}$. This shows that $\mathbf{C}\in\mathcal{Q}_{\E,\Delta t}$ for $\mathfrak{E}=\Delta E_{\rho_0}$ and hence for all $\E\geq\Delta E_{\rho_0}$: that is, the sets $\mathcal{Q}_{\E,\Delta t}$ describe the correlations in the scenario of Fig.~\ref{fig:setup} with or without shared randomness $\lambda$, obtainable from average prepared states $\rho_0$ with energy uncertainty $\Delta E_{\rho_0}\leq\E$.

For $\E\Delta t\rightarrow0$, the states have overlap $\gamma\rightarrow1$, and so are indistinguishable by the measurement device. In this case, the outputs $b$ must be independent of $x$; i.e.\ the correlations are on the line $C_0=C_1$. However, for increasing $\E\Delta t$, the overlap decreases such that the measurement device can at least partially distinguish the two states -- therefore a larger set of correlations are available, as shown in Figure~\ref{fig:correlations}. For $\E\Delta t\geq\pi/2$, the states may be perfectly distinguishable, and therefore all correlations are possible. 

Given the above, let us now summarise the assumption that our semi-DI framework imposes:

\textbf{Assumption.} \emph{In every single run of the experiment, on input $x\in\{0,1\}$, the preparation device P emits a system in some state $\rho_x$, unentangled with the measurement device M. Input $x=1$ means waiting some known amount of time $\Delta t>0$ as compared to $x=0$. In this time interval, the system evolves unitarily according to some Hamiltonian $\hat H$. While $\hat H$ may be unknown, we assume an upper bound $\mathfrak{E}$ on the energy uncertainty $\Delta E$ of the system in state $\rho_0$ (and thus $\rho_1$), as given by $\hat H$.}

Note that we make no assumptions about the unknown convex components $\rho_x^\lambda$ of~(\ref{EqWithLambda}), or about the measurement device. Furthermore, we show in the ``Open system evolution'' section below how the assumption of unitary evolution can be substantially relaxed. If $\hat H$ is known, then $\Delta E$ can be determined by, for example, measuring $\hat H$ (and thus $\hat H^2$), or by determining the standard deviation of measurements of $\hat H$ over many runs. Moreover, since energy is a conserved quantity, it is conceivable to come up with protocols that estimate the energy (and its standard deviation) indirectly even without knowing $\hat H$. For example, P may be powered by an energy source, and we may read off the total ``energy bill'' after a large number of runs. We leave an analysis of this idea to future work.

\subsubsection*{Classical correlations}

Suppose that an adversary has complete knowledge of the variable $\lambda$ in the decomposition~(\ref{EqWithLambda}), and knows the value of $x$. Under what conditions does this allow the adversary to predict the measurement outcome $b$ perfectly, for both $x\in\{0,1\}$? To answer this question, we have to consider the energy uncertainty of the states $\rho_0^\lambda$ if we know that $\Delta E_{\rho_0}\leq\E$. In Appendix~\ref{concavity}, we show that the standard deviation $\Delta E$ is \textit{concave}, and so
\[
\Delta E_{\rho_0} \geq \sum_\lambda p(\lambda)\Delta E_{\rho_0^\lambda}.
\]
That is, while some $\rho_0^\lambda$ may have energy uncertainty strictly larger than $\E$, the average energy uncertainty is still upper-bounded by $\E$.

Now suppose that the eavesdropper, knowing $\lambda$, can perfectly predict the output of $M$ conditioned on $x$. This means that the associated correlations $\mathbf{C}^\lambda=(C_0^\lambda,C_1^\lambda)$ must be in the set of deterministic correlations, $\{\pm 1\}\times\{\pm 1\}$, and that the energy bound must be satisfied on average. Thus, the average correlation $\mathbf{C}=\sum_\lambda p(\lambda)\mathbf{C}^\lambda$ must be contained in the following set, the \textit{classical max-average set}, defined similarly as in~\cite{VanHimbeeck}:
\begin{align}
\overline{\mathcal{C}}_{\E,\Delta t}=& \left\{ \mathbf{C}=\sum_\lambda p(\lambda)\mathbf{C}^\lambda\,\,\middle|\,\, \mathbf{C}^\lambda\in\mathcal{Q}_{\E^\lambda,\Delta t}, \right. \nonumber\\ 
& \quad \left. \mathbf{C}^\lambda\in\{\pm 1\}\times\{\pm 1\},\sum_\lambda p(\lambda)\E^\lambda\leq \E\right\}.
\label{DefMaxAverage}
\end{align}
Here, we allow probabilistic mixtures to contain some ``fast'' (i.e. large $\Delta E$) components, provided the maximum on the average value is respected.

To characterise this set, we consider two possibilities for each $\lambda$; either $\E^\lambda<\frac{\pi}{2\Delta t}$, in which case the only deterministic correlations are those represented by the line $|C^\lambda_0-C^\lambda_1|=0$; or $\E^\lambda\geq\frac{\pi}{2\Delta t}$, in which case all correlations are possible, i.e. $|C^\lambda_0-C^\lambda_1|\leq2$. We collect the variables according to $\Lambda_1=\{\lambda_1:\E^{\lambda_1}<\frac{\pi}{2\Delta t}\}$, which occurs with probability $\sum_{\lambda_1\in\Lambda_1}p(\lambda_1)=:p_1$ and $\Lambda_2=\{\lambda_2:\E^{\lambda_2}\geq\frac{\pi}{2\Delta t}\}$, which occurs with probability $\sum_{\lambda_2\in\Lambda_2}p(\lambda_2)=:p_2=1-p_1$, and where $\Lambda=\Lambda_1\dot{\cup}\Lambda_2$. First, consider the case that both $p_1$ and $p_2$ are non-zero. The correlations are bounded according to
\begin{align*}
    |C_0-C_1|\leq\sum_{\lambda\in\Lambda} p(\lambda)|C^\lambda_0-C^\lambda_1|\leq 2p_2.
\end{align*}
We have
\begin{align*}
&\sum_{\lambda_1\in\Lambda_1}p(\lambda_1)\E^{\lambda_1}+\sum_{\lambda_2\in\Lambda_2}p(\lambda_2)\E^{\lambda_2}\leq \E \\
\Rightarrow\,&p_1\sum_{\lambda_1\in\Lambda_1}\underbrace{\frac{p(\lambda_1)}{p_1}}_{:=q_{\lambda_1}}\E^{\lambda_1} + p_2\sum_{\lambda_2\in\Lambda_2}\underbrace{\frac{p(\lambda_2)}{p_2}}_{:=q_{\lambda_2}}\E^{\lambda_2}\leq \E \\
\Rightarrow\,&p_1\E_1+p_2\E_2\leq \E. 
\end{align*}
In the last line, we have defined $\E_1:=\sum_{\lambda_1\in\Lambda_1}q_{\lambda_1}\E^{\lambda_1}$ and $\E_2:=\sum_{\lambda_2\in\Lambda_2}q_{\lambda_2}\E^{\lambda_2}$ (the averages of the energy uncertainties below and above $\frac{\pi}{2\Delta t}$ respectively), for which $\sum_{\lambda_1\in\Lambda_1}q_{\lambda_1}=1$ and $\sum_{\lambda_2\in\Lambda_2}q_{\lambda_2}=1$.
From this, we can then calculate a bound on $p_2$:
\begin{align}
    p_2&\leq \frac{\E-\E_1}{\E_2-\E_1}=\frac{\delta}{\epsilon+\delta}\nonumber
\end{align}
where, by construction, $\E_1<\E<\E_2$, so we can write $\E_1=\E-\delta$ and $\E_2=\E+\epsilon$, with $0<\delta\leq \E$ and $\epsilon\geq\frac{\pi}{2\Delta t}-\E\geq0$. This is clearly maximised by taking the lower bound $\epsilon=\frac{\pi}{2\Delta t} -\E$ (i.e.\ $\E_2=\frac{\pi}{2\Delta t}$, the minimal $\E_2$ for which all correlations are possible). Then, the remaining fraction is maximised by taking the upper bound $\delta=\E$ (i.e.\ $\E_1=0$). Therefore:
\begin{align}
    p_2\leq\frac{2\E\Delta t}{\pi}.
    \label{p2}
\end{align}
So far, we have assumed that $p_1,p_2\neq 0$. But if $p_2=0$, then this bound is trivially true. Furthermore, if $p_1=0$, then $\E=\E_2\geq\frac\pi{2\Delta t}$, and so $\frac{2\E\Delta t}\pi \geq 1 = p_2$. Hence~(\ref{p2}) is true in all cases.

Therefore we have the following bound on the correlations, where the lower case is the one of interest for the max-average assumption:
\begin{equation}\label{classical_max_av}
|C_0-C_1|\leq 
\begin{cases} 
    2 & \text{if}\,\, \E\geq\frac{\pi}{2\Delta t}, \\
    \frac{4\E\Delta t}{\pi} & \text{if}\,\, \E<\frac{\pi}{2\Delta t}.
\end{cases}
\end{equation}
Conversely, suppose some correlation $\bm{k}=(k_0,k_1)$ satisfies inequality (\ref{classical_max_av}). It is geometrically clear that this must be in $\overline{\mathcal{C}}_{\E,\Delta t}$, following the same line of reasoning as in \cite[Appendix D]{RotationBoxes1}. Suppose $k_0< k_1$, then one can draw a line through $\bm{k}$, from one corner $(-1,1)$ to the diagonal line $|C_0-C_1|=0$, intersecting at some $\bm{C}=(C,C)$. Therefore, this correlation can be written as $\bm{k}=\kappa(-1,1)+(1-\kappa)\bm{C}$, for some $\kappa\in[0,1]$. From this, we know $\kappa=\frac{1}{2}(k_1-k_0)\leq\frac{2\E\Delta t}{\pi}$. Define $\Lambda=\{1,2,3\}$, $\mathbf{C}^1=(-1,1)$, $\mathbf{C}^2=(-1,-1)$, and $\mathbf{C}^3=(+1,+1)$. We have just shown that $\mathbf{k}$ can be written as a convex combination of these three correlations, where the weight $p(1)$ of $\mathbf{C}^1$ is at most $\frac{2\E\Delta t}\pi$. Moreover, $\mathbf{C}^\lambda\in\mathcal{Q}_{\E^\lambda,\Delta t}$ for $\E^1=\frac\pi{2\Delta t}$ and $\E^2=\E^3=0$. Thus, the energy constraint in definition~(\ref{DefMaxAverage}) is satisfied:
\[
\sum_\lambda p(\lambda)\E^\lambda = p(1)\E^1 \leq\frac{2\E\Delta t}\pi\cdot\frac\pi {2\Delta t}=\E.
\]
This shows that $\mathbf{k}\in \overline{\mathcal{C}}_{\E,\Delta t}$. The case $k_0\geq k_1$ can be treated analogously swapping the extremal point $(1,-1)$ for $(-1,1)$. Thus, Eq.~(\ref{classical_max_av}) characterises the $\overline{\mathcal{C}}_{E,\Delta t}$ precisely.

\begin{figure}[t]
\centering 
\includegraphics[trim=0 0 400 0,clip, width=0.9\columnwidth]{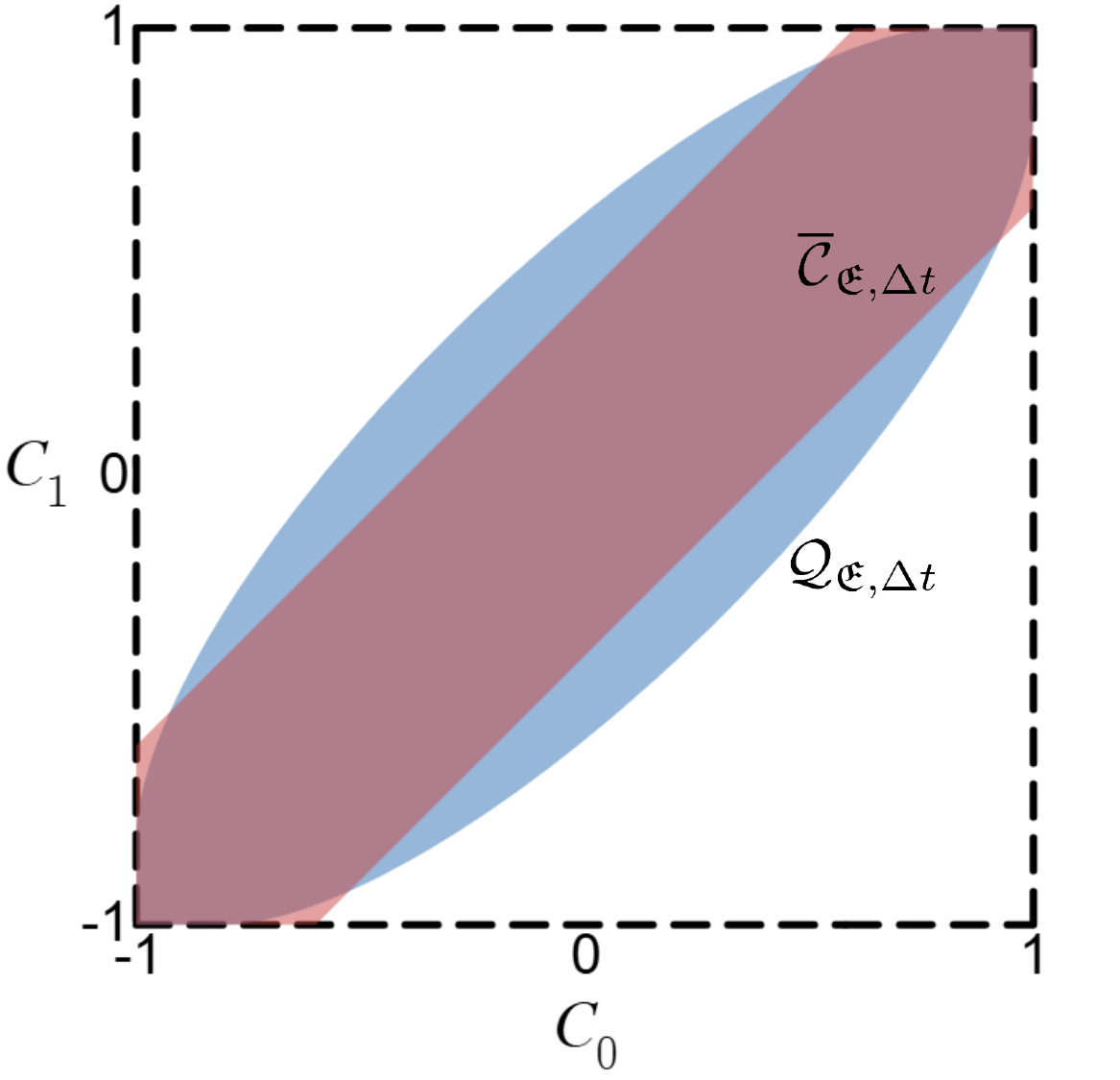} 
\caption{Set of quantum correlations $\mathcal{Q}_{\E,\Delta t}$ (blue) and classical max-average correlations $\mathcal{\bar C}_{\E,\Delta t}$ (red) for $\E\Delta t=0.314$ (and thus $\gamma=0.951$ according to Eq.~(\ref{EqGamma})).}
\label{fig:max-average}
\end{figure}

For an illustration of the classical max-average set $\mathcal{\bar C}_{\E,\Delta t}$ see Fig.~\ref{fig:max-average}. Since a fraction $p_2$ of ``fast'' (large $\Delta E$) components are allowed, and the max-average constraint implies $p_2 \leq 2E\Delta t/\pi$, we get the linear wedge $|C_0 - C_1| \leq 4E\Delta t/\pi$. Interestingly, it is \textit{not} in general a strict subset of the quantum set: for small $\E\Delta t$, there are correlations that can be modelled classically under the max-average assumption that are not predicted by quantum theory under the stricter $\E$ constraint. This constraint is stricter because
\[
   \Delta E_{\rho_0}\leq\mathfrak{E}\quad\begin{array}{c} \Rightarrow \\ \not\Leftarrow \end{array} \quad\sum_\lambda p(\lambda)\Delta E_{\rho_0^\lambda}\leq\mathfrak{E}.
\]
The observation that $\overline{\mathcal{C}}_{\mathfrak{E},\Delta t}\not\subset\mathcal{Q}_{\mathfrak{E},\Delta t}$ does therefore not signal any interesting insights into the quantum-classical distinction, but is simply a consequence of the mathematical fact that the energy uncertainty is not linear, but concave.

Hence, if we have an upper bound $\E$ on the energy uncertainty of the initial state, and if our scenario generates some correlation $\mathbf{C}\in \mathcal{Q}_{\E,\Delta t}\cap \mathcal{\bar C}_{\E,\Delta t}$, then it is possible that this correlation comes from an ensemble of deterministic correlations, which makes the outcome $b$ potentially predictable by an adversary. But importantly for our protocol, for all $0<\E\Delta t<\frac{\pi}{2}$, there are correlations consistent with $\mathcal{Q}_{\E,\Delta t}$ that are outside of $\overline{\mathcal{C}}_{\E,\Delta t}$. That is, such correlations are inconsistent with a deterministic model, even when we can only measure the \textit{average} value of the energy uncertainty, rather than assuming its validity for every value of $\lambda$ separately. This allows us to certify randomness, the quantification of which we discuss in Section~\ref{randomness}.

\subsubsection*{Open system evolution}

So far we have formulated our semi-DI assumption for closed system communication, under unitary evolution. However, we can consider a natural extension to open systems, in which the inevitable environmental effects are taken into account. In particular, \cite{Shiraishi} derive a speed limit for open systems $S$ coupled to general environments $E$ via an arbitrary coupling Hamiltonian
\[
   \hat{H}=\hat{H}_S+\hat{H}_{SE}+\hat{H}_E=\hat{H}'_S+\hat{H}_E,
\]
where $\hat{H}'_S=\hat{H}_S+\hat{H}_{SE}$ describes the Hamiltonian of $S$ together with its coupling to the environment. If $\rho_0$ is the quantum state of $S$ at time $0$ and $\rho_1$ at time $\Delta t$, then they show that
\begin{align}
    \Delta t\geq\frac{\arccos \mathcal{F}(\rho_0,\rho_1)}{\langle \Delta E\rangle_{\Delta t}},
    \label{EqMTNew}
\end{align}
where $\mathcal{F}(\rho_0,\rho_1)={\rm tr}\sqrt{\rho_0^{1/2}\rho_1\rho_0^{1/2}}$ is the Uhlmann fidelity, and $\langle\Delta E\rangle_{\Delta t}=\frac 1 {\Delta t}\int_0^{\Delta t}dt\,\Delta E(t)$ is the time average of the energy uncertainty $\Delta E(t)=\sqrt{\langle {\hat{H'}}_S^2\rangle_{\rho(t)}-(\langle {\hat{H}'}_S\rangle_{\rho(t)})^2}$, with $\rho(t)$ the reduced state of $S$ at time $t$. For pure states $\rho_k=|\psi_k\rangle\langle\psi_k|$, the fidelity is $\mathcal{F}(\rho_0,\rho_1)=|\langle\psi_0|\psi_1\rangle|$, and for $\hat{H}={\hat{H}'}_S$, this bound reduces to the Mandelstam-Tamm bound~(\ref{QSLpure}).

In cases where $S$ is ``small'' compared to its environment $E$, the authors of~\cite{Shiraishi} show that their bound~(\ref{EqMTNew}) is a significant improvement of the original Mandelstam-Tamm bound (which, indeed, becomes trivial if the environment is very large). Moreover, as we will now show, our results above continue to hold unchanged if the assumption $\Delta E_{\rho_0}\leq\E$ is replaced by an assumption $\langle \Delta E\rangle_{\Delta t}\leq\E$ for the time-averaged open-system energy uncertainty. In some sense, rather than having to assume something about the system's energy uncertainty \textit{within the whole universe}, we now only have to assume a bound on its energy uncertainty \textit{within its vicinity of influence}.

To see this, we will first show that $\mathcal{F}(\rho_0,\rho_1)\geq\gamma$ implies that Eq.~(\ref{EqGamma}) holds for all measurement procedures, i.e.\ all $-\mathbb{1}\leq M\leq\mathbb{1}$ that generate the correlations $C_x={\rm tr}(\rho_x M)$. This can be seen as follows. According to Uhlmann's theorem, there exist purifications $|\psi_i\rangle_{SA}$ of $\rho_i$ ($i=0,1$), where $A$ is some ancillary system, such that $\mathcal{F}(\rho_0,\rho_1)=|\langle\psi_0|\psi_1\rangle_{SA}|$. But then, Eq.~(\ref{EqGamma}) applies whenever $C_x=\langle\psi_x|M'_{SA}|\psi_x\rangle_{SA}$, where $-\mathbb{1}\leq M'_{SA}\leq\mathbb{1}$. But this is true in particular for $M'_{SA}=M_S\otimes\mathbf{1}_A$, where $C_x=\langle\psi_x|M_S\otimes\mathbf{1}_A|\psi_x\rangle_{SA}={\rm tr}(\rho_x M)$.

From Eq.~(\ref{EqMTNew}), we have $\mathcal{F}(\rho_0,\rho_1)\geq\gamma$, where
\[
   \gamma=\left\{
      \begin{array}{cl}
         \cos\left(\langle\Delta E\rangle_{\Delta t}\Delta t\right) & \mbox{if }\left|\langle\Delta E\rangle_{\Delta t}\Delta t\right|<\frac\pi 2 \\
         0 & \mbox{otherwise}.
      \end{array}
   \right.
\]
Thus the possible quantum correlations obtainable under open-system evolution as described above under the assumption $\langle\Delta E\rangle_{\Delta t}\leq\E$ are exactly given by $\mathcal{Q}_{\E,\Delta t}$. All further arguments, including the randomness certification results below, continue to hold without changes.

For a possibly even tighter quantum bound, Ref.~\cite{Taddei} proves a general constraint for arbitrary physical processes in terms of the quantum Fisher information $\mathcal{F}_{Q}(t)$. Whilst their inequality serves as a tighter bound in general on the state transformation, the quantum Fisher information is less easy to directly determine for time-dependent evolution~\cite{Shiraishi,Taddei}. Therefore the formulation of~\cite{Shiraishi} in terms of the energy uncertainty $\langle E_S\rangle_\tau$ is more readily suitable for our operationally motivated scenario.

\section{Randomness certification}\label{randomness}

For correlations $\bm{C}\in\mathcal{Q}_{\E,\Delta t}\backslash \overline{\mathcal{C}}_{\E,\Delta t}$, there is no classical model that could reproduce correlations predicted by quantum theory, even assuming the energy constraint is only respected on average. Therefore, observing such correlations certifies randomness in that they have no deterministic description, and can be used for the generation of random numbers. The amount of certified randomness can be quantified adapting a method of \cite{VanHimbeeck2}. 

Imagine that an adversary (Eve) wishes to guess the output of $b$. Like the experimenter (Alice), she may have knowledge of the inputs $x$, but (unlike Alice) she may also have access to additional information about the scenario. We may even consider that she knows all additional physical parameters, characterised by $\lambda$, that determine the behaviour of the device. We could imagine in some situations that some physical parameter, of which Eve knows the value, is correlated with the outcome $b$, allowing her a predictive advantage over the experimenter. We wish to prove for our scenario that, irrespective of knowing any such parameters, the outcome of the experiment is random, even to Eve. 

\begin{figure}[t]
\centering 
\includegraphics[trim=50 0 40 0,clip, width=1.0\columnwidth]{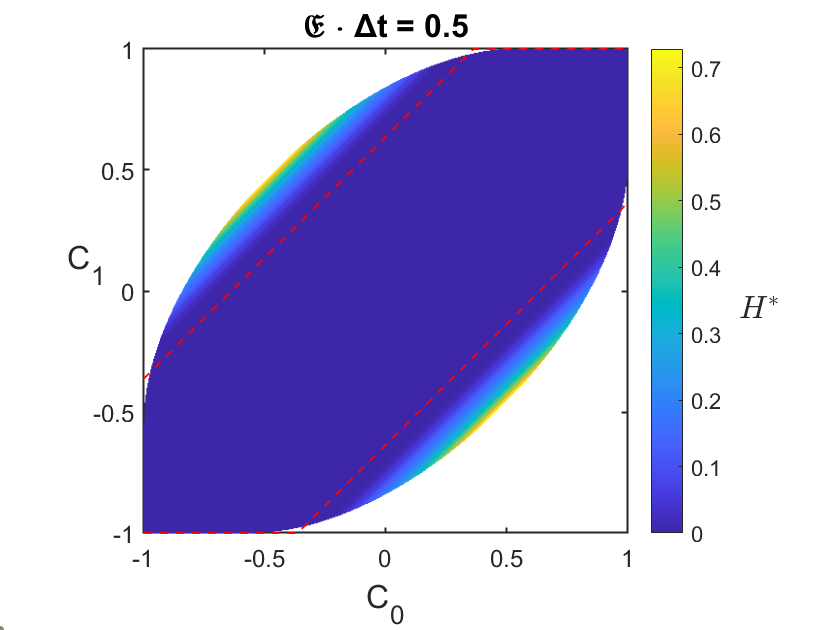} 
\caption{Numerical certified lower bounds on the entropy $H^\star$ under the assumption $\Delta E \leq \E$ with $\E\Delta t=0.5$ (in units where $\hbar=1$), obtained using the discretisation parameters $L=20$, $M=5$, and $N=S=5000$ (See Appendix~\ref{sec-algorithm}). Points with $H^\star > 0$ correspond to correlations that allow for certified randomness extraction. As expected, $H^\star=0$ (only) on the red-rimmed classical max-average set $\mathcal{\bar C}_{\E,\Delta t}$. With this discretisation, the largest certified lower bound in the plot is $H^\star\geq 0.7284$, which occurs at $(C_0,C_1) = (\pm \sin(\E\Delta t), \mp \sin(\E\Delta t))$. A more exhaustive run (Appendix~\ref{sec-maxH}) with $N=S=20000$ yields the certified lower bound $H^\star\geq 0.8113$. This value may still not be quite tight, and finer discretisation could yield marginally higher $H^\star$. See Appendix~\ref{appendix-minentropy} for all details on the methodology we use to evaluate~(\ref{opt_prob}).}
\label{fig:min-entropy}
\end{figure}

For Eve, the scenario is characterised by the ensemble $\{p(\lambda),(\bm{C}^\lambda,\E^\lambda)\}$. The correlations as viewed by the experimenter can be more precisely characterised according to Eve's knowledge, by $\bm{C}=\sum_\lambda p(\lambda)\bm{C}^\lambda$. Nevertheless, the experimenter has already checked that the energy uncertainty is on average bounded by $\E$, therefore the constraint $\sum_\lambda p(\lambda)\E^\lambda \leq \E$ applies, which follows from the concavity of variances (see Appendix \ref{concavity}). For Alice to be sure that Eve cannot predict the outcome $b$ reliably, she can quantify the randomness via e.g.\ the conditional Shannon entropy $H(B|X,\Lambda)=-\sum_{b,x,\lambda}p(b,x,\lambda)\log_2 p(b|x,\lambda)$. This quantifies the difficulty for Eve to predict $b$, given $x$ and $\lambda$. Provided the inputs are independent of $\lambda$, the conditional entropy can be written as 
\[
H(B|X,\Lambda)=\sum_\lambda p(\lambda) H(\mathbf{C}^\lambda),
\]
where 
\[H(\bm{C}):=-\frac{1}{2}\sum_{b,x} \frac{1+b C_x}{2} \log_2 \frac {1+b C_x}{2}.
\]
This can be determined by the optimisation problem $H(B|X,\Lambda)\geq H^\star$, where
\begin{subequations}\label{opt_prob}
\begin{eqnarray}
H^\star&=&\min_{\{p(\lambda),\bm{C}^\lambda,\E^\lambda\}} \sum_\lambda p(\lambda) H(\bm{C}^\lambda)\label{opt_prob_a}\\
&& \mbox{s.t. } \sum_\lambda p(\lambda)\mathbf{C}^\lambda=\bm{C},\label{opt_prob_b} \\
&& \sum_{\lambda}p(\lambda)\E^\lambda= \E, \label{opt_prob_c} \\
&& \mathbf{C}^\lambda\in\mathcal{Q}_{\E^\lambda,\Delta t}.\label{opt_prob_d}
\end{eqnarray}
\end{subequations}
As in~\cite[Eq.\ 26]{VanHimbeeck2}, we take the condition in~(\ref{opt_prob_c}) to be equality without loss of generality; see Appendix~\ref{equality}. In contrast to their setting, however, our feasible set defined by the constraint~(\ref{opt_prob_d}) is non-convex when ranging over all possible $\E^\lambda$.

In Appendix~\ref{appendix-minentropy}, we give a detailed analysis of this optimisation problem, and a concrete algorithm that provides certified lower bounds to $H^\star$. Even though our domain of optimisation is not convex, we show that the Lagrange dual of the problem, as in~\cite{VanHimbeeck2}, can be formulated and reproduces $H^\star$ exactly. We then provide several insights into the optimisation problem: for example, we show that $H^\star$ is convex, that its value at the midpoints of the lines $\{(C_0,C_1)^\top\in\mathcal{Q}_{\mathfrak{E},\Delta t}\,\,|\,\, C_0-C_1={\rm const.}\}$ lower-bounds the value everywhere else on the lines, and we further constrain the domain of optimisation in the dual problem. As apparent in Fig.~\ref{fig:min-entropy}, $H^\star$ attains the same values at the points $(C_0,C_1)$, $(C_1,C_0)$, $(-C_0,-C_1)$ and $(-C_1,-C_0)$. Finally, we provide a discretisation of the dual problem, still yielding certified lower bounds to $H^\star$, by analytically bounding the error arising from the discretisation. These analytic results ensure that our algorithm has acceptable runtime and provides rigorous lower bounds to $H^\star$, converging to $H^\star$ from below in the limit of finer and finer discretisation.

In Figure~\ref{fig:min-entropy}, we plot the resulting estimates of $H^\star$ over the space of correlation pairs, for a fixed average energy uncertainty constraint $\mathfrak{E}\Delta t=1/2$. Any value $H^\star>0$ can be used to extract a semi-device-independent certificate of randomness. As expected, points lying within the classical average set $\bar{\mathcal{C}}$ yield $H^\star=0$ (and vice versa), confirming that no randomness can be certified against an adversary in that region, but a non-zero amount of randomness can be certified everywhere outside of it.

\section{Implementation with coherent states}
\label{SecExperimental}

\begin{figure*}[t]
\centering 
\includegraphics[trim=0 280 70 0,clip, width=1.0\linewidth]{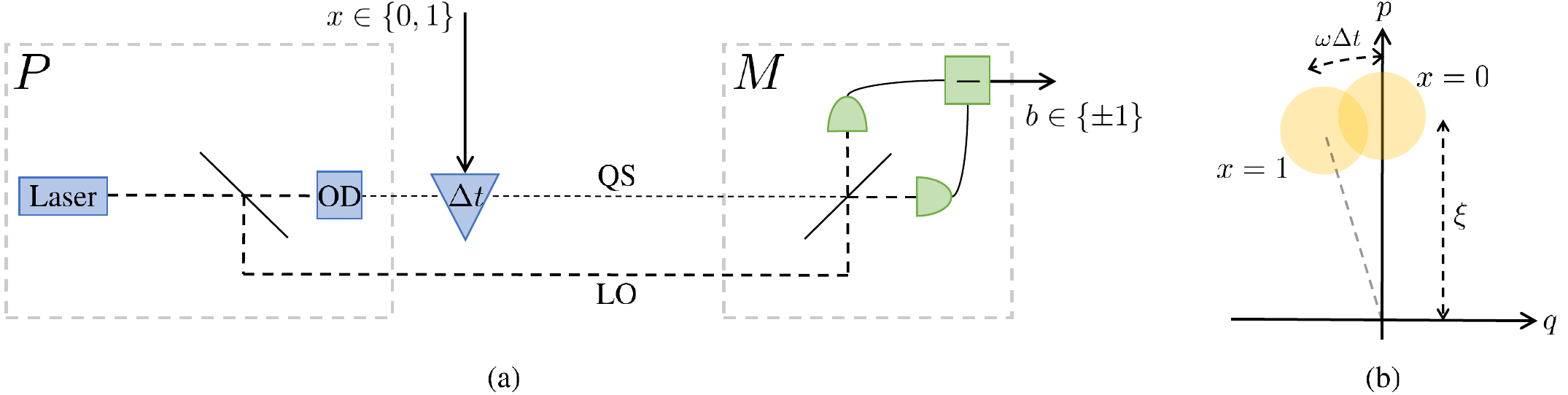}
\caption{(a) Diagram of proposed experimental setup; technically similar to, but conceptually distinct from the BPSK implementation of \cite{VanHimbeeck2}. The preparation device $P$ consists of a monochromatic laser emitting a coherent pulse, which is sent through a beam splitter for homodyne detection. The transmitted beam (the quantum signal, QS) is attenuated by an optical density filter (OD) and then, for the input $x=1$, delayed by some time $\Delta t$.
This can be achieved using an optical wedge, whose position and thickness depends on the input $x$.
By causing a time delay, the wedge introduces a corresponding phase shift.
At the measurement device $M$, the signal is interfered with the reflected signal (the local oscillator, LO), which acts as a phase reference, and is measured. The quadrature $q$ is then measured, the sign of which determines the output $b$. (b) Diagram of the phase space of the coherent state, showing the relative phase $\omega\Delta t$ between the two states that may arrive at $M$. For $x=0$, the state $\ket{\alpha}=\ket{i\xi}$ is centred on the y-axis (in the rotating frame of reference of the local oscillator), evolving counter-clockwise with period $2\pi/\omega$. For $x=1$, the state is given by $\ket{\alpha(t=\Delta t)}=\ket{i\xi e^{-i\omega\Delta t}}$.}
\label{fig:coherent_setup}
\end{figure*}

In this section, we show that our protocol can in principle be performed with coherent states of a single harmonic oscillator, and we sketch a quantum-optical implementation.

Consider a situation in which the initial state is given by $\ket{\psi_0}=\ket{\alpha}$ a single-mode coherent state, described in its Fock basis by
\begin{equation}\label{eq:coherentstate}
    \ket{\alpha}=e^{-\frac{|\alpha|^2}{2}}\sum^\infty_{n=0}\frac{\alpha^n}{\sqrt{n!}}\ket{n}.
\end{equation}
The state evolves in time as a harmonic oscillator $\hat{H}=\hbar\omega(\hat{a}^\dag \hat{a}+\frac{1}{2})$, and we use the quantum optics convention where $\hat a=\frac 1 {\sqrt{2}}(\hat q+i\hat p)$ and $[\hat a,\hat a^\dagger]=\mathbf{1}$ such that $\hat H=\frac{\hbar\omega}2 (\hat q^2+\hat p^2)$ (an analogous but more cumbersome calculation for the mechanical oscillator $\hat H=\frac{\hat p^2}{2m}+\frac 1 2 m\omega^2 \hat q^2$ leads to the same result for the correlations $\mathbf{C}=(C_0,C_1)$ as below). The coherent state evolves in time by rotating in phase space,
\begin{equation}
    e^{-i\hat{H}t}\ket{\alpha}=
    e^{-i\omega t/2}|\alpha(t)\rangle,
\end{equation}
where $\alpha(t)=e^{-i\omega t}\alpha$. Up to an irrelevant global phase $\theta\in\mathbb{R}$, its representation in the quadrature $q$ basis is given by
\[
   \langle q|\alpha\rangle=e^{i\theta}\pi^{-1/4}\exp\left[ -\frac 1 2 \left(q-\sqrt{2}\,{\rm Re}(\alpha)\right)^ 2\right],
\]
where $\ket{q}$ is an eigenstate of the quadrature operator $\hat{q}=\frac 1 {\sqrt{2}}(\hat{a}+\hat{a}^\dag)$.
This expression shows that the coherent state is a Gaussian wavepacket centered at $\sqrt{2}\,{\rm Re}(\alpha)$ in the quadrature $q$.

At the measurement device $M$, we follow a protocol similar to the Binary Phase Shift Keying (BPSK) example of \cite[Section 2.3.1]{VanHimbeeck}. In particular, $M$ implements a quadrature measurement and assigns the binary output $b=\textrm{sign}(q)$, i.e.
\begin{equation}
    b=\begin{cases}
        +1 & \textrm{if}\,q>0,\\
        -1 & \textrm{if}\,q<0.
    \end{cases}
\end{equation}
The measurement outcome is sampled from
\begin{align*}
    |\braket{q}{\alpha(t)}|^2&= \pi^{-1/2} \exp\left[-\left(\strut q-\sqrt{2}\, {\rm Re}(\alpha e^{-i\omega t})\right)^2\right].
\end{align*}
That is, $b=b(t)$ is time-dependent. This leaves room for there to be interesting correlations from our scenario. In particular, the correlations are given by
\begin{eqnarray}
    C(t)&\equiv&\mathbb{E}[b(t)]=\int\limits_{-\infty}^{\infty}\mathrm{sign}(q)|\langle q|\alpha(t)\rangle|^2 dq  \nonumber\\
    &=& \mathrm{erf}(\sqrt{2}\mathrm{Re}(\alpha e^{-i\omega t})),
    \label{eq:correlationfunction}
\end{eqnarray}
where the final line uses the well-known result $\mathbb{E}[\textrm{sign}(x)]=\mathrm{erf}(\mu/\sqrt{2}\sigma)$ for a Gaussian distribution centred at $\mu$ with standard deviation $\sigma$ (here, we have $\sigma=1/\sqrt{2}$). Then the correlations $(C_0,C_1)$ are given by
\begin{align}
    C_0&=C(t=0)=\mathrm{erf}(\sqrt{2}\mathrm{Re}(\alpha)),\nonumber\\
    C_1&=C(t=\Delta t)=\mathrm{erf}(\sqrt{2}\mathrm{Re}(\alpha e^{-i\omega\Delta t})).\label{eqCorr01}
\end{align}

The energy expectation is given by
\begin{equation*}
\langle\hat{H}\rangle_{\alpha(t)}=\hbar\omega\left(|\alpha(t)|^2+\frac{1}{2}\right)=\hbar\omega\left(|\alpha|^2+\frac{1}{2}\right),
\end{equation*}
and the squared energy:
\begin{equation*}
\langle\hat{H}^2\rangle_{\alpha(t)}=\hbar^2\omega^2\left(|\alpha(t)|^4+2|\alpha(t)|^2+\frac{1}{4}\right).
\end{equation*}
The energy uncertainty is thus given by
\begin{align*}
\Delta E_{\alpha(t)}&=\sqrt{\langle\hat{H}^2\rangle-\langle\hat{H}\rangle^2}=\hbar\omega|\alpha(t)|=\hbar\omega|\alpha|.
\end{align*}

In order to realise interesting quantum correlations that certify randomness, suppose we begin with an initial state $\ket{\alpha}=\ket{i\xi}$ where $\xi\in\mathbb{R}$. In this case, the energy uncertainty is given by $\Delta E_\alpha=\hbar\omega\xi$. For the correlation function, from Eq.~(\ref{eqCorr01}), we get 
\begin{align}
    C_0&=0,\label{C0_xi}\\
    C_1&=\textrm{erf}(\sqrt{2}\xi\sin(\omega \Delta t)).\label{C1_xi}
\end{align}

To summarise, in order to obtain interesting correlations that certify randomness from coherent states, we want values of $\bm{C}=(0,\textrm{erf}(\sqrt{2}\xi\sin\omega \Delta t))$ that are in the set $\mathcal{Q}_{\E,\Delta t}$ but outside $\overline{\mathcal{C}}_{\E,\Delta t}$, where $\hbar\omega\xi\leq\E$. From the characterisation of the max-average classical set of equation (\ref{classical_max_av}), with units $\hbar$ put back in, we get the following condition for non-zero entropy:
\begin{equation}\label{coherent_randomness}
\left|\textrm{erf}\left(\sqrt{2}\xi\sin(\omega \Delta t)\right)\right|>\frac{4\E \Delta t}{\pi\hbar},
\end{equation}
for $\E\Delta t/\hbar<\pi/2$ and $\hbar\omega\xi\leq\E$.

\begin{figure}[t]
\centering 
\includegraphics[trim=00 0 260 0,clip, width=0.9\columnwidth]{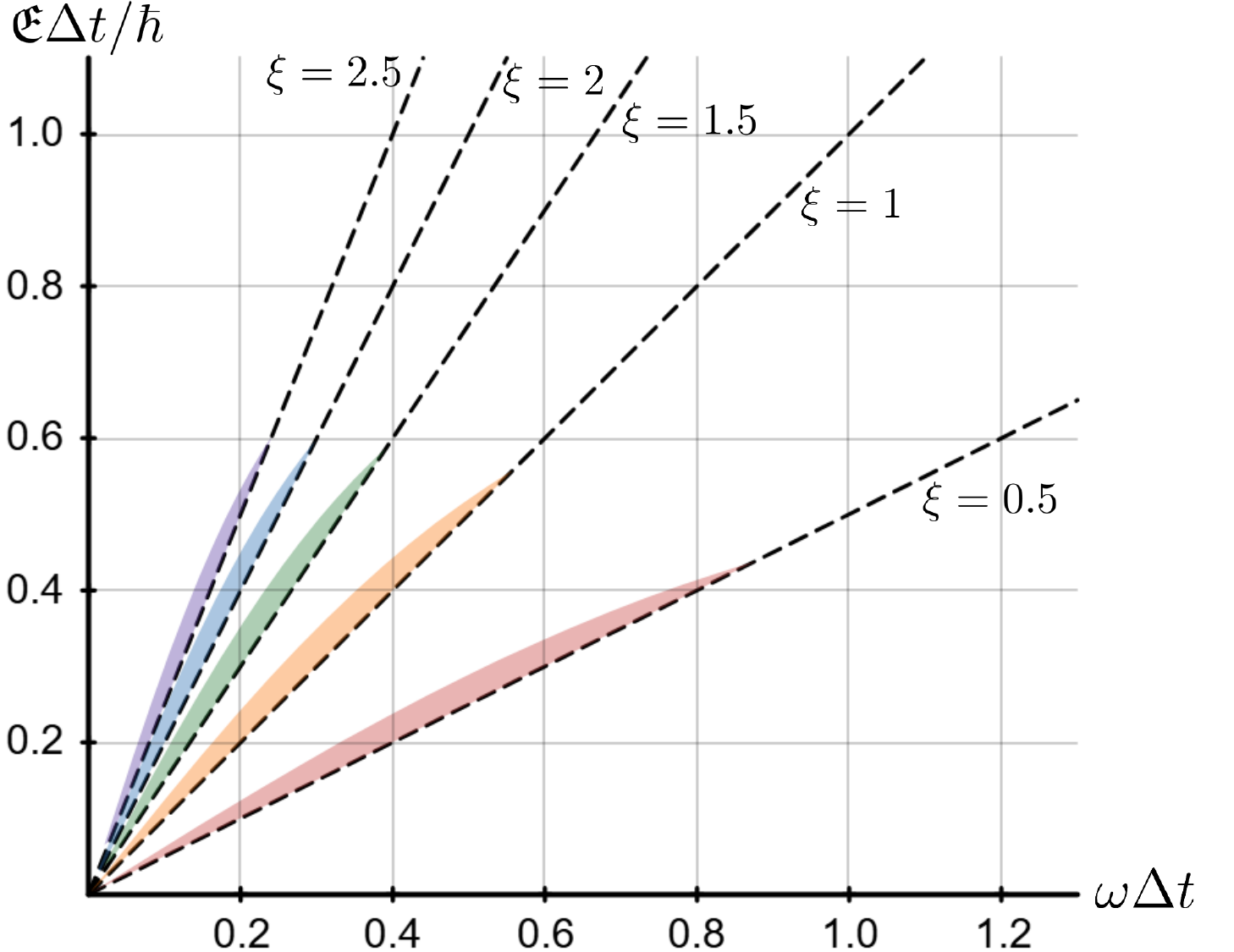} 
\caption{Regions for which non-zero randomness can be certified according to Eq.~(\ref{coherent_randomness}), for $\xi\in\{0.5,1,1.5,2,2.5\}$ (red, yellow, green, blue, purple), where $\xi=|\alpha|$ is related to the average photon number $\langle n\rangle=\xi^2$ in the coherent state $|\alpha\rangle$. Dashed lines represent the lines $\hbar\omega\xi=\E$.}
\label{fig:coherent_state}
\end{figure}

As Fig.~\ref{fig:coherent_state} shows, there are indeed choices of frequency $\omega$, time delay $\Delta t$ and of assumed upper bounds $\E$ to the energy uncertainty such that a non-zero amount of randomness can be certified with a standard coherent state of a simple harmonic oscillator. An example is for $\hbar\omega\xi=\E=0.5$, $\omega\Delta t=0.4$ (i.e.\ the point $(0.4,0.2)$ in Fig.~\ref{fig:coherent_state}), for which we obtain $C_1=0.303$. This places $\bm{C}\in\mathcal{Q}_{\E,\Delta t}\backslash \overline{\mathcal{C}}_{\E,\Delta t}$, and yields a certified entropy $H^\star=0.0242$.

In Fig.~\ref{fig:coherent_setup}, we sketch a possible quantum-optical implementation, see the figure caption for a description. Note that the local oscillator can in principle be replaced by a second laser, and then only the upper path corresponds to the physical system that is prepared and measured. Alternatively, as in e.g.~Implementation 1 of \cite{BohrBrask2017}, the laser could simply be initiated at one of two possible times. Since the implementation of the time delay relies on the specific physics of light propagation in materials, it does not accomplish the device-independence of the anticipated ``trusted operation on an untrusted device'' depicted in Fig.~\ref{fig:setup}. However, assuming that a photon state has been prepared, and that the optical element works as desired, it admits the certification of randomness independently of the actually prepared photon state (and of the actually performed measurement).

\section{Discussion \& outlook}\label{SecOutlook}

In this work we have demonstrated the applicability of the quantum speed limit for performing tasks in quantum information -- namely, for generating certifiably secure random numbers. We have described a simple prepare-and-measure scenario, where the inputs correspond to two time-displaced preparations of a transmitted system. This communicated system is constrained only in terms of some upper bound on its energy uncertainty. We have characterised the possible quantum correlations of the scenario, for pure state, mixed state and open systems, and shown that quantum theory predicts statistics that are incompatible with any deterministic explanation -- even for an adversary who has knowledge of the inputs and who has access to extra information about the communication between devices. Moreover, we have provided a numerical estimate for the amount of certifiable entropy $H^\star$ under our energy and time assumptions, and sketched an experimental implementation involving coherent states that yields non-zero certified entropy. The certification of genuine randomness from coherent states bears a resemblance to an observation by e.g.~Tsirelson~\cite{Tsirelson,Zaw} indicating some form of nonclassical behavior of the simple harmonic oscillator.

Harnessing the QSL for information-theoretic protocols builds on existing literature in which uncertainty principles have been viewed as a hallmark of nonclassicality (e.g.~\cite{Berta,Winter,Tomamichel,Buscemi,Catani}); in our case, the dynamical form of quantum uncertainty gives rise to the gap between quantum and classical sets of correlations. The notion of characterising correlations via time-separated measurements has also been explored in the recent preprint~\cite{Manos}, in the so-called ``extrapolation problem''~\cite{Araujo}.

Our protocol contributes to ongoing efforts to replace the traditional dimensionality assumption by physically better motivated alternatives, building on the work of e.g.~\cite{BohrBrask2017,Tebyanian2021,RochiCarceller2024,Ioannou2022,Tavakoli2021,tavakoli2020informationally,TZC2022,Pauwels,VanHimbeeck,VanHimbeeck2}. The security of our protocol is grounded in reliable upper bounds on the energy uncertainty and the time delay between preparations. The former could be verified ``from the outside'', in a similar spirit as proposed by~\cite{VanHimbeeck}; by performing many tests on the state $\rho_x$ emitted by $P$, one could do a statistical analysis to determine the energy uncertainty -- i.e. the second moment of the measured probability distribution. The latter is imposed manifestly by the nature of the experiment; the device time-displaces the two possible states by $\Delta t$, by virtue of implementing a time delay during one of the two preparations. Ideally, we would consider that the experimenter themselves waits some time between preparation times, but the nano-timescale prohibits a fully operational implementation of this assumption. Nevertheless, we can conceive of this as a trusted super-operation on an otherwise uncharacterised preparation box.

The consideration of semi-DI protocols via trusted operations in spacetime on untrusted devices was initiated in~\cite{RotationBoxes1}. There, the input for a prepare-and-measure scenario is given by some rotation of the preparation device around a fixed axis by some angle $\alpha$, while assuming an upper bound to the spin $J$ of the transmitted physical system. For small enough angles, the gap between quantum and classical (deterministic) sets of correlations can be used to generate secure random numbers. Moreover, it has been shown that the set of quantum correlations in this setup can be recovered even without assuming quantum theory. In a similar spirit to the ``spacetimes boxes'' framework of~\cite{RotationBoxes1,RotationBoxes2}, there may be scope to extend the results of this paper to a theory-independent setting. There has been some interesting work in this direction by other authors~\cite{Giannelli}, who derive a speed limit from purely information-theoretic principles. This could motivate the generalisation of the results of this paper beyond quantum theory, or even within a framework guided only by spatiotemporal considerations. If this were possible, one could look to ground quantum information protocols, such as the certification of randomness, on assumptions about time translation symmetry alone. 
\\

\section*{Acknowledgments}
We acknowledge support from the Austrian Science Fund (FWF) via project P 33730-N. Furthermore, this research was funded in part by the Austrian Science Fund (FWF) 10.55776/PAT2839723, and it was funded by the European Union - NextGenerationEU. This research was supported in part by Perimeter Institute for Theoretical Physics. Research at Perimeter Institute is supported by the Government of Canada through the Department of Innovation, Science, and Economic Development, and by the Province of Ontario through the Ministry of Colleges and Universities.

\restoretoc

\section*{Appendix}

\setcounter{section}{0}
\renewcommand\thesection{\Alph{section}}
\renewcommand{\thesubsection}{\roman{subsection}}

\section{Mixed states}\label{appendix-mixed}

We now consider the set of quantum correlations for mixed states, which we prove is equal to that of pure states. For $\E\geq 0$ and $\Delta t\geq 0$, we define the following:
\begin{align}
    \mathcal{Q}'_{\E, \Delta t}:=\Big\{& (C_0,C_1) \, \big| \, C_x=\tr[M\rho_x], -\mathbb{1}\leq M\leq\mathbb{1},  \nonumber\\
    &\exists \hat{H} \,\text{s.\ t.\ }\rho_1=U_{\Delta t}^\dagger\rho_0 U_{\Delta t},\, \Delta E_{\rho_0}\leq \E\Big\},
\end{align}
which we show to be equal to $\mathcal{Q}_{\E, \Delta t}$. To do so, we consider the purification of mixed states, using ancilla systems, and show that the purification procedure does not change the energy variance. In particular, consider a mixed state $\rho^A_0$ acting on $\mathcal{H}_A$, which is diagonalised as $\rho^A_0=\sum_{k^A}\tilde{c}_{k}\ket{k^A}\bra{k^A}$. To characterise its purification, we introduce a fictitious system $B$ with Hilbert space $\mathcal{H}_B$ such that $\mathrm{dim}(\mathcal{H}_A)=\mathrm{dim}(\mathcal{H}_B)$. This defines a pure state $\ket{\psi_{AB}}=\sum_k \sqrt{\tilde{c}_k}\ket{k^Ak^B}$, evolving under the Hamiltonian $\hat{H}_{AB}:=\hat{H}_A+{\hat0}_B$ via $U_{\Delta t}\otimes\mathbb{1}_B$.
This embedding does not affect the energy variance. In particular,
\[
\langle \hat{H}^2_{AB}\rangle_{\psi_{AB}}=\bra{\psi_{AB}}\hat{H}^2_A\otimes\id_B \ket{\psi_{AB}}=\tr[\rho_A \hat{H}^2_A]= \langle \hat{H}^2_{A}\rangle_{\rho_{A}},
\]
and likewise for $\langle \hat{H}_{AB}\rangle_{\psi_{AB}}$. Therefore, $\Delta E_{\psi_{AB}}=\Delta E_{\rho_{A}}$.
We can think of the communication of a mixed state $\rho_A$ as being realised by the communication of a pure state $\ket{\psi}_{AB}$ (which is therefore already contained within the characterisation $\mathcal{Q}_{\E,\Delta t}$), but such that the measurement device does not ``listen'' to the subsystem $B$. This means that we embed the POVM $\{M_+,M_-\}$ on $A$ via $\{M_+\otimes\mathbb{1}_B,M_-\otimes\mathbb{1}_B\}$ on $AB$, reproducing the original correlation via a pure state on the larger system $AB$. 

Accordingly, we have proven the equality
\begin{equation}
    \mathcal{Q}'_{\E, \Delta t}=\mathcal{Q}_{\E, \Delta t}.
\end{equation}
That is, the set of quantum correlations is unchanged by including mixed states. This also implies that $\mathcal{Q}_{\E,\Delta t}$ is convex.
\\

\section{Quantum model for all correlations of (\ref{quantum-pure})}
\label{quantum-model}

Having shown that all correlations in $\mathcal{Q}_{\E,\Delta t}$ are constrained by inequality (\ref{quantum-pure}), we now prove the tightness of this bound by providing a quantum model for all correlations that satisfy (\ref{quantum-pure}), that is:
\begin{equation*}
\frac{1}{2}\left(\sqrt{1+C_0}\sqrt{1+C_1}+\sqrt{1-C_0}\sqrt{1-C_1}\right) \geq \gamma.
\end{equation*}
It is useful to switch description from correlations $\bm{C}=(C_0,C_1)$ to probabilities $\bm{P}^+=(P^+_0,P^+_{1})$, where probabilities $P^+_x:=P(+1|x)$ are related to the correlations by the bijective affine transformation $\bm{P}^+=(\bm{C}+1)/2$. In probability space, we can write down a description for the inequality above in terms of the extremal points $(0,0)$, $(1,1)$, along with the following curves $p_1$ and $p_2$ parametrised by $\tau$:
\begin{align}
    p_1(\tau)&=\left(\cos^2({\E\tau}),\cos^2({\E(\tau+\Delta t)})\right),\,\, \tau\in\mathcal{I}_1, \label{parametrisation1}\\
    p_2(\tau)&=\left(\cos^2({\E\tau}),\cos^2({\E(\tau-\Delta t)})\right),\,\, \tau\in\mathcal{I}_2,\label{parametrisation2}
\end{align}
where $\mathcal{I}_1=[0,\frac{\pi}{2\E}-\Delta t]$ and $\mathcal{I}_2=[\Delta t,\frac{\pi}{2\E}]$.

We will now provide a quantum model for all of these extremal points. First, the points $(0,0)$ and $(1,1)$ are given by constant probability distributions, and so can be trivially modelled. For example, the latter can be reproduced by the state $\ket{\psi_0}=\ket{\psi_1}=\ket{E}$, for which $\Delta E=0$, and the measurement operators $M_+=\ketbra{E}$ and $M_-=\mathbb{1}-M_+$. 

Next we take the curve given by $p_1(\tau)$, which can be modelled (for instance) by the state 
\[\ket{\psi_0}=\frac{\ket{0}+\ket{2\E}}{\sqrt{2}}\] 
for some Hamiltonian that has $0$ and $2\E$ among its energy eigenvalues,
which has an energy uncertainty of $\Delta E=\E$. This state evolves in a time $\Delta t$ to the following:
\begin{align*}
    \ket{\psi_1}&=U_{\Delta t}\left(\frac{\ket{0}+\ket{2\E}}{\sqrt{2}}\right)\\
    &=\frac{1}{\sqrt{2}}\left(
    \ket{0}+e^{-2i{\E\Delta t}}\ket{2\E}\right) \\
    &=\frac{1}{\sqrt{2}}e^{-i \E\Delta t}\left(e^{i{\E\Delta t}}\ket{0}+e^{-i{\E\Delta t}}\ket{2\E}\right).
\end{align*}
Then we define the measurement operators $M_+=U^\dagger_\tau\ketbra{\psi_0}U_\tau$ and $M_-=\mathbb{1}-M_+$, for which a simple calculation shows that we achieve the required probabilities: $P_0^+=\langle\psi_0|M_+|\psi_0\rangle=\cos^2(\E\tau)$ and $P_1^+=\langle\psi_1|M_+|\psi_1\rangle=\cos^2(\E(\tau+\Delta t))$. The curve $p_2(\tau)$ can be modelled similarly, using the same state but the measurement operators $M_+=U^\dagger_{-\tau}\ketbra{\psi_0}U_{-\tau}$ and $M_-=\mathbb{1}-M_+$. 

Having modelled all extreme points, we then use that $\mathcal{Q}_{\E,\Delta t}$ is convex to conclude that all correlations that obey inequality (\ref{quantum-pure}) have a quantum model, and thus it precisely characterises the quantum set $\mathcal{Q}_{E, \Delta t}$.

\section{Concavity of $\Delta E$}
\label{concavity}

To show concavity of $\Delta E_\rho$ in $\rho=\lambda\rho_1+(1-\lambda)\rho_2$, first we note convexity of $\langle \hat{H}\rangle^2_\rho$:
\begin{align*}
\langle \hat{H}\rangle_{\rho}^2 & =(\lambda\langle \hat{H}\rangle_{\rho_1}+(1-\lambda)\langle \hat{H}\rangle_{\rho_2})^2\\
&\leq\lambda\langle \hat{H}\rangle^2_{\rho_1}+(1-\lambda)\langle \hat{H}\rangle^2_{\rho_2},
\end{align*}
due to the convexity of $(\cdot)^2$.
We also have the equality
\begin{align*}
\lambda\langle \hat{H}^2\rangle_{\rho_1}+(1-\lambda)\langle \hat{H}^2\rangle_{\rho_2}=\langle \hat{H}^2\rangle_\rho.
\end{align*}
Together with the concavity of the square root function, these show concavity of $\Delta E_\rho$ in $\rho$:
\begin{align*}
\Delta E_\rho &= \sqrt{\langle \hat{H}^2\rangle_\rho-\langle \hat{H}\rangle^2_\rho} \\
&\geq \sqrt{\lambda\langle \hat{H}^2\rangle_{\rho_1}+(1\!-\!\lambda)\langle \hat{H}^2\rangle_{\rho_2} - \lambda\langle \hat{H}\rangle^2_{\rho_1}-
(1\!-\!\lambda)\langle \hat{H}\rangle^2_{\rho_2}} \\
&= \sqrt{\lambda(\langle \hat{H}^2\rangle_{\rho_1}-\langle \hat{H}\rangle^2_{\rho_1})+(1-\lambda)(\langle \hat{H}^2\rangle_{\rho_2}-\langle \hat{H}\rangle^2_{\rho_2}) } \\
&= \sqrt{\lambda(\Delta E_{\rho_1})^2+(1-\lambda)(\Delta E_{\rho_2})^2 } \\
&\geq \lambda\Delta E_{\rho_1}+(1-\lambda)\Delta E_{\rho_2}\, .
\end{align*}
This extends in an obvious way to more general convex combinations $\rho=\sum_\lambda p(\lambda)\rho_\lambda$.

\section{Equality~(\ref{opt_prob_c}) of the optimisation problem}\label{equality}

We need to show the following: if $\{p(\lambda),\mathbf{C}^\lambda,\E^\lambda\}$ satisfies (\ref{opt_prob_b}), (\ref{opt_prob_c'}) and (\ref{opt_prob_d}), where
\begin{equation}
\sum_\lambda p(\lambda)\E^\lambda\leq\E,
\tag{14c'}\label{opt_prob_c'}
\end{equation}
then there are ${\E'}^\lambda\geq\E^\lambda$ such that $\{p(\lambda),\mathbf{C}^\lambda,{E'}^\lambda\}$ satisfies (\ref{opt_prob_b}), (\ref{opt_prob_c}) and (\ref{opt_prob_d}). Consequently, we can replace (\ref{opt_prob_c'}) by (\ref{opt_prob_c}) in the optimisation problem.

This can be seen as follows. Clearly, $\sum_\lambda p(\lambda){\E'}^\lambda$ is increasing in ${\E'}^\lambda$, and it can in fact be made arbitrarily large. Thus, we can pick the ${\E'}^\lambda$ such that $\sum_\lambda p(\lambda){\E'}^\lambda=\E$, i.e.\ (\ref{opt_prob_c}) is satisfied. Since (\ref{opt_prob_b}) depends only on the correlations $\mathbf{C}^\lambda$, it is unchanged. Finally, we have
\[
\E\leq\E'\Rightarrow \mathcal{Q}_{\E,\Delta t}\subseteq \mathcal{Q}_{\E',\Delta t},
\]
and so (\ref{opt_prob_d}) implies that $\mathbf{C}^\lambda\in\mathcal{Q}_{{\E'}^\lambda,\Delta t}$.

\section{Optimisation framework for Bounding the Entropy $H^\star$}
\label{appendix-minentropy}

We are interested in bounding the amount of randomness (entropy $H^\star$) that can be certified in our prepare-and-measure scenario from given observed correlations $\bm{C}=(C_0,C_1)$, under an average energy uncertainty constraint $\Delta E_{\rho_0}\leq\mathfrak{E}$ and evolution time $\Delta t$. The goal is to minimise the conditional entropy $H(B|X,\Lambda)$ over all compatible hidden-variable decompositions. Formally, the optimisation problem can be expressed as:
\begin{eqnarray}\label{eq:primal}
\begin{aligned}
    H^\star := \min_{\{p(\lambda),\, \bm{C}^\lambda,\, \mathfrak{E}^\lambda\}} \quad & \sum_\lambda p(\lambda) H(\bm{C}^\lambda) \\
    \text{subject to} \quad & \sum_\lambda p(\lambda) \bm{C}^\lambda = \bm{C}, \\
                            & \sum_\lambda p(\lambda) \mathfrak{E}^\lambda = \mathfrak{E}, \\
& \bm{C}^\lambda \in \mathcal{Q}_{\mathfrak{E}^\lambda, \Delta t},
\end{aligned}
\end{eqnarray}
where $H(\bm{C}) = \sum_x p(x) h_{\rm bin}(C_x)$, using the binary entropy function $h_{\rm bin}(C_x)=-\sum\limits_{b}\frac{1+b C_x}{2}\log_2 \frac{1+b C_x}{2}$. Here we assume $p(x)=1/2$ for both $x$, i.e.\ there is equal a priori probability for both inputs.

In this section, we develop an algorithm that computes an approximation to $H^\star=H^\star(\mathbf{C,\mathfrak{E}})$, i.e.\ to the amount of certifiable randomness for every possible observed correlation $\mathbf{C}$ and energy uncertainty $\mathfrak{E}$. The approximation $H^\star(L,M,N,S)$ will depend on four discretisation parameters $L,M,N,S\in\mathbb{N}$, such that, in the limit of these parameters turning to infinity (in a suitable order), this approximation will converge to the correct value $H^\star$. Moreover, we will prove that the approximation is always a lower bound, i.e.\ $H^\star(L,M,N,S)\leq H^\star$ for all $L,M,N,S$ and all $\mathbf{C}$ and $\mathfrak{E}$. This will be done in the following steps:
\begin{itemize}
    \item In \textbf{Subsection~\ref{SubsecDualFormulation}}, we construct a dual formulation of the optimisation problem~(\ref{eq:primal}), and show that it maintains the same optimal value $H^\star$. The main advantage of doing so is to get rid of the optimisation over an unbounded number of ensembles $\{p(\lambda),\mathbf{C}^\lambda,\mathfrak{E}^\lambda\}$. Its feasible points $(t,\bm t)$ define affine functions $t+\bm t\cdot(\bm C,\mathfrak E)$ lying below the entropy function $H(\bm C)$ on the entire domain $\mathcal Q$.
    \item In \textbf{Subsection~\ref{SubsecPropertiesDual}}, we prove several properties of the dual optimisation problem and its solution. In addition to the fundamental insights, these are used for significant speed-ups of the algorithm that is developed in the following subsections.
    \item The dual~\eqref{eq:dual} involves an infinite family of nonlinear entropy constraints, which makes it highly non-trivial to solve for the exact optimum. In \textbf{Subsection~\ref{SubsecDiscretisation}}, we deal with this issue by discretising the problem, yielding a finite optimisation domain.
    We approximate the dual by restricting the outer maximisation to a finite grid of dual variables (\textit{e.g.,} $\bm{t}\in I^3_{L,M}$). Since it only restricts the maximisation, the resulting value is a lower bound to the true optimum (Lemma~\ref{LemFirstDiscr}).
    \item However, the dual objective function involves $t(\bm{t})$, defined through inner minimisations. Discretising these minimisation problems requires additional care, since naively replacing the continuous domain by a finite subset can overestimate the minimum value, thereby invalidating the bound. To avoid this, at each discretisation step, we subtract explicit worst-case error terms, ensuring that any possible overestimation is compensated so that the computed quantity $t_{L,N,S}(\bm t)$ is guaranteed to satisfy $t_{L,N,S}(\bm t)\le t(\bm t)$ (Lemma~\ref{LemDiscrE} adds the correction term associated with discretising the energy uncertainty domain $\frak{E}' \in \left[0,\pi/2\right]$, while Lemma~\ref{lemmaCertified_lb} adds the one for discretising the set $\mathcal{Q}_{\frak{E}',1}$).
    \item As a result, every approximation step either restricts a maximisation or underestimates a minimisation by construction, ensuring that the final output $H^\star(L,M,N,S)$ always satisfies $H^\star(L,M,N,S)\leq H^\star$ as desired.
    \item In the limit of increasingly fine discretisation, these error terms vanish and  $H^\star(L,M,N,S)$ converges to the exact value $H^\star$ (see Lemma~\ref{lemmaCertified_lb}).
    \item In \textbf{Subsection~\ref{sec-algorithm}}, we describe the concrete algorithm that we use to implement this method. We introduce a significant speed-up for producing the plot of Figure~\ref{fig:min-entropy} in \textbf{Subsection~\ref{SubsecDiagonal}}, and give some comments on a special case in \textbf{Subsection~\ref{sec-maxH}}.
\end{itemize}

\subsection{Dual formulation}
\label{SubsecDualFormulation}

Finding an optimal solution for~(\ref{eq:primal}) requires, in principle, optimising over all possible ensembles $\{p(\lambda), \mathfrak{E}^\lambda, \bm{C}^\lambda\}$. Since the number of hidden variables $\lambda \in \Lambda$ is unbounded, this problem is generally intractable in its full form. To address this challenge, we follow a strategy similar to the one presented in \cite{VanHimbeeck2} by formulating the dual problem. This reformulation allows us to obtain safe lower bounds on the entropy $H^\star_{\textrm{dual}}\leq H^\star$, even when the primal problem is not directly solvable. Moreover, we demonstrate that our problem exhibits the strong duality property, ensuring that the optimal dual solution coincides with the optimal primal solution, \textit{i.e.,} $H^\star_{\textrm{dual}} = H^\star$.

Before deriving the dual, it is convenient to first express the primal problem as part of a broader class of non-linear constrained optimisation problems. Let $f$ be a continuous function defined over the set of feasible solutions $\mathcal{Q} \subset \mathbb{R}^{\textrm{dim} (\mathcal{Q})}$. The optimisation problem can then be expressed as:
\begin{eqnarray}
\begin{aligned}
    f^\star(\bm{x}_0) = \min_{\{\bm{x}^\lambda,p(\lambda)\}} \quad & \sum_\lambda p(\lambda) f(\bm{x}^\lambda) \\
    \text{subject to} \quad & \sum_\lambda p(\lambda) \bm{x}^\lambda = \bm{x}_0, \\
 & \bm{x}^\lambda \in \mathcal{Q}, \\
& p(\lambda) \in \mathcal{P}(\Lambda),
\end{aligned}
\label{eq:gral_primal}
\end{eqnarray}
where $\mathcal{P}(\lambda)$ denotes the set of all probability distributions over the $\lambda$. For our specific problem of Eq.~\eqref{eq:primal}, the feasible set $\mathcal{Q} \subset \mathbb{R}^3$ contains all possible correlations $\bm{C}^\lambda \in \mathcal{Q}_{\mathfrak{E}^{\lambda},\Delta t}$ for all energy uncertainties within $0 \leq \mathfrak{E}^\lambda \leq \pi/(2\Delta t)$. Therefore, the optimisation variables can be identified as $\bm{x}^\lambda = (\bm{C}^\lambda,\mathfrak{E}^\lambda) \in \mathcal{Q}$, while the given observed data and constraint are $\bm{x}_0 = (\bm{C},\mathfrak{E})$. 

\begin{figure}[b]
\centering 
\includegraphics[width=0.8
\columnwidth]{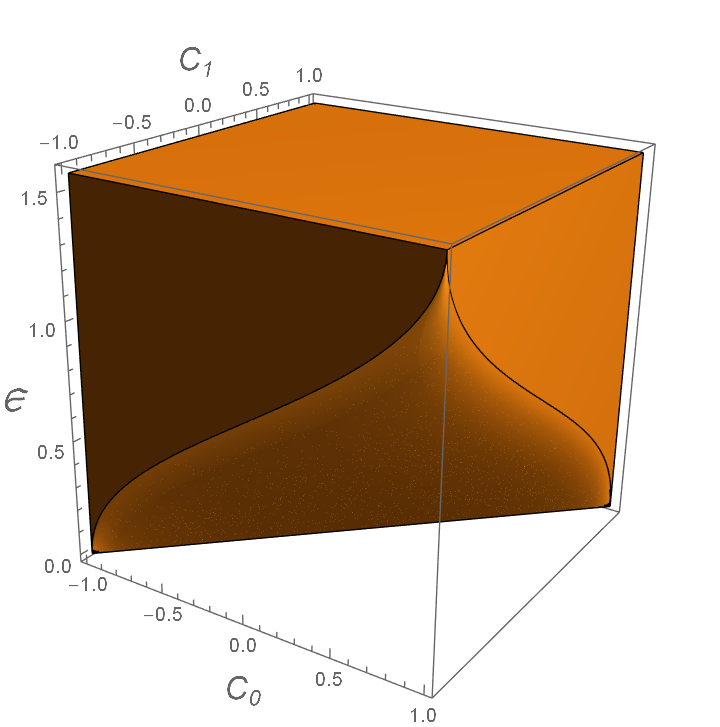}
\caption{The non-convex feasible set $\mathcal{Q}$ for the optimisation problem~\ref{eq:primal}, representing the set of allowed correlations and energy uncertainties used as optimisation variables. Specifically, the elements of the set are of the form $(C_0^\lambda,C_1^\lambda,\mathfrak{E}^\lambda)$.}
\label{fig:feasibleSet}
\end{figure}

In contrast to the case in \cite{VanHimbeeck2}, our feasible set $\mathcal{Q}$ is non-convex, as illustrated in Figure~\ref{fig:feasibleSet}. Despite this non-convexity, it can be demonstrated that the dual formulation of the problem maintains strong duality, ensuring that the optimal solution obtained from the dual problem coincides with the optimal primal solution.

To derive the dual formulation of the optimisation problem, we begin by defining the Lagrangian:
{\small
\begin{eqnarray}
\mathcal{L}(\bm{x}_0,\{p(\lambda),\bm{x}^\lambda\}, \bm{t}) = \sum_\lambda p(\lambda)f(\bm{x}^\lambda) + \bm{t} \cdot \left( \bm{x}_0 -  \sum_\lambda  p(\lambda)\bm{x}^\lambda \right), \nonumber
\end{eqnarray}}
where $\bm{t}\in\mathbb{R}^m$ are the Lagrange multipliers (or \textit{dual variables}) associated to each constraint in the primal problem. In our case~\eqref{eq:primal}, we have $m=3$ corresponding to the constraints on the observed values $C_0, C_1$ and $\mathfrak{E}$. 

The dual function $g : \mathbb{R}^m \to \mathbb{R}$ is then defined as:
{\small
\begin{eqnarray}
 g(\bm{x}_0,\bm{t}) &=& \inf\limits_{\{p(\lambda),\bm{x}^\lambda\}} \mathcal{L}(\bm{x}_0,\{p(\lambda),\bm{x}^\lambda\},\bm{t}) \\
 &=& \inf\limits_{\{p(\lambda),\bm{x}^\lambda\}} \left\{ \sum_\lambda p(\lambda)f(\bm{x}^\lambda) + \bm{t} \cdot \left( \bm{x}_0 -  \sum_\lambda  p(\lambda)\bm{x}^\lambda \right)\right\} \nonumber\\
  &=& \bm{t}\cdot\bm{x}_0 + \inf\limits_{\{p(\lambda),\bm{x}^\lambda\}} \left\{  \sum\limits_\lambda p(\lambda) \left(f(\bm{x}^\lambda) - \bm{t}\cdot \bm{x}^\lambda\right)\right\} . \nonumber
\end{eqnarray}}
The dual function $g$ is concave by construction, as it is a pointwise infimum over affine functions. Next, let $t\in\mathbb{R}$ be the infimum $t := \inf\limits_{\{p(\lambda),\bm{x}^\lambda\}} \left\{  \sum\limits_\lambda p(\lambda) \left(f(\bm{x}^\lambda) - \bm{t}\cdot \bm{x}^\lambda\right)\right\}$, so that we can express the dual function simply as $ g(\bm{x}_0,\bm{t}) = t + \bm{t}\cdot\bm{x}_0$. 

This dual Lagrange function satisfies the \textit{weak duality} property, yielding lower bounds on the optimal value:
\[g(\bm{x}_0,\bm{t}) \leq f^\star(\bm{x}_0),\]
where $f^\star(\bm{x}_0)=\min_{\{p(\lambda),\bm{x}^\lambda\}}\sum_\lambda p(\lambda)f(\bm{x}^\lambda)$ is the optimal value of the primal problem. To see that weak duality is satisfied, consider any feasible ensemble $\{p(\lambda),\bm{x}^\lambda\}$ such that $\sum_\lambda p(\lambda) \bm{x}^\lambda = \bm{x}_0$. Then:
\begin{eqnarray}
g(\bm{x}_0,\bm{t}) &\leq & \bm{t}\cdot \bm{x}_0 + \sum\limits_\lambda p(\lambda)(f(\bm{x}^\lambda)-\bm{t}\cdot\bm{x}^\lambda) \\
&=&  \bm{t}\cdot\bm{x}_0 + \sum\limits_\lambda p(\lambda)f(\bm{x}^\lambda) -  \bm{t}\cdot\sum\limits_\lambda p(\lambda)\bm{x}^\lambda \nonumber\\
&= & \sum\limits_\lambda p(\lambda)f(\bm{x}^\lambda) . \nonumber
\end{eqnarray}
Since this holds for any feasible ensemble, we conclude $g(\bm{x}_0,\bm{t})\leq f^\star(\bm{x}_0)$. Furthermore, note that for all $\bm{x}\in\mathcal{Q}$ one has $t+\bm{t}\cdot\bm{x}\leq f(\bm{x})$. 

The inequality
\[t+\bm{t}\cdot\bm{x} \leq f(\bm{x}) \quad \forall\bm{x}\in\mathcal{Q}\]
then defines a global affine underestimator of the function $f$. Therefore, the dual problem corresponding to the primal~\eqref{eq:gral_primal} can be expressed as
\begin{eqnarray}
\begin{aligned}
    \sup_{t, \bm{t}} \quad & t + \bm{t} \cdot \bm{x}_0 \\
    \text{subject to} \quad & t + \bm{t}\cdot \bm{x} \leq f(\bm{x}) \quad \forall \bm{x} \in \mathcal{Q}.
\end{aligned}
\label{eq:dual_gral}
\end{eqnarray}
In our prepare-and-measure scenario, each variable $\bm{x}=(C_0',C_1',\mathfrak{E}')\in\mathbb{R}^3$ represents a possible correlation pair and corresponding energy uncertainty. The observed quantities $\bm{x}_0 = (C_0,C_1, \mathfrak{E})$ are given and fixed. The feasible set is: 
\[
\mathcal{Q}:=\{ (C_0',C_1',\mathfrak{E}') \,\,|\,\,  (C_0',C_1') \in \mathcal{Q}_{\mathfrak{E}',\Delta t}, 0 \leq \mathfrak{E}' \leq \frac{\pi}{2 \Delta t} \},
\]
where recall that $\mathcal{Q}_{\mathfrak{E}',\Delta t}\subset \left[-1,1\right]^2$ denotes the set of physically allowed correlations for a given energy uncertainty $\mathfrak{E}'$ and evolution time $\Delta t$. 
Finally, the primal objective function $f(\bm{x})$ is the entropy function $f(\bm{x})=H(\bm{C}')$ defined above. 

Thus, the dual formulation of our problem becomes:
\begin{eqnarray}
\begin{aligned}
    H^\star_{\textrm{dual}} = \sup_{(t, \bm{t})\in\mathbb{R}\times\mathbb{R}^3} \quad & t + \bm{t} \cdot \left(\begin{array}{c} \bm{C}  \\  \mathfrak{E} \end{array}\right)\\
    \text{subject to} \quad & t + \bm{t}\cdot \left(\begin{array}{c} \bm{C}'  \\  \mathfrak{E}' \end{array}\right) \leq H(\bm{C}') \quad \forall \left(\begin{array}{c} \bm{C}'  \\  \mathfrak{E}' \end{array}\right) \in \mathcal{Q}.
\end{aligned}
\label{eq:dual}
\end{eqnarray}
The objective function $H$ is concave. It is easy to see that this implies that both in the primal and in the dual problem, we can replace the domain $\mathcal{Q}$ by its convex hull ${\rm conv}(Q)$ (in~(\ref{eq:primal}), to do so, note that the last line is equivalent to $\left(\begin{array}{c}\bm{C}^\lambda \\ \mathfrak{E}^\lambda\end{array}\right)\in\mathcal{Q}$). Since this is a convex set, Proposition 18 in~\cite{VanHimbeeck2} implies strong duality, i.e.\ $H^\star=H^\star_{\rm dual}$ for all $\mathbf{C}$ and $\E$.

\subsection{Properties of the dual problem and of $H^\star$}
\label{SubsecPropertiesDual}

In this subsection, we prove some results that will in the following allow us to reduce the domain of optimisation in~(\ref{eq:dual}): rather than considering all $(t,\bm{t})\in \mathbb{R}\times\mathbb{R}^3$, we will only have to consider a subregion with a certain combination of signs, as specified in Lemma~\ref{lemma_tvec}. Furthermore, in the other lemmas, we show that $H^\star$ has certain symmetry and convexity properties that will make the computation of Figure~\ref{fig:min-entropy} much more efficient.

In this and the following subsections, let us choose units such that $\Delta t=1$, i.e.\ the relevant values of $\mathfrak{E}$ are in $[0,\frac\pi 2]$.

\begin{figure}[t]
\centering 
\includegraphics[trim=0 0 400 0,clip,width=0.9\columnwidth]{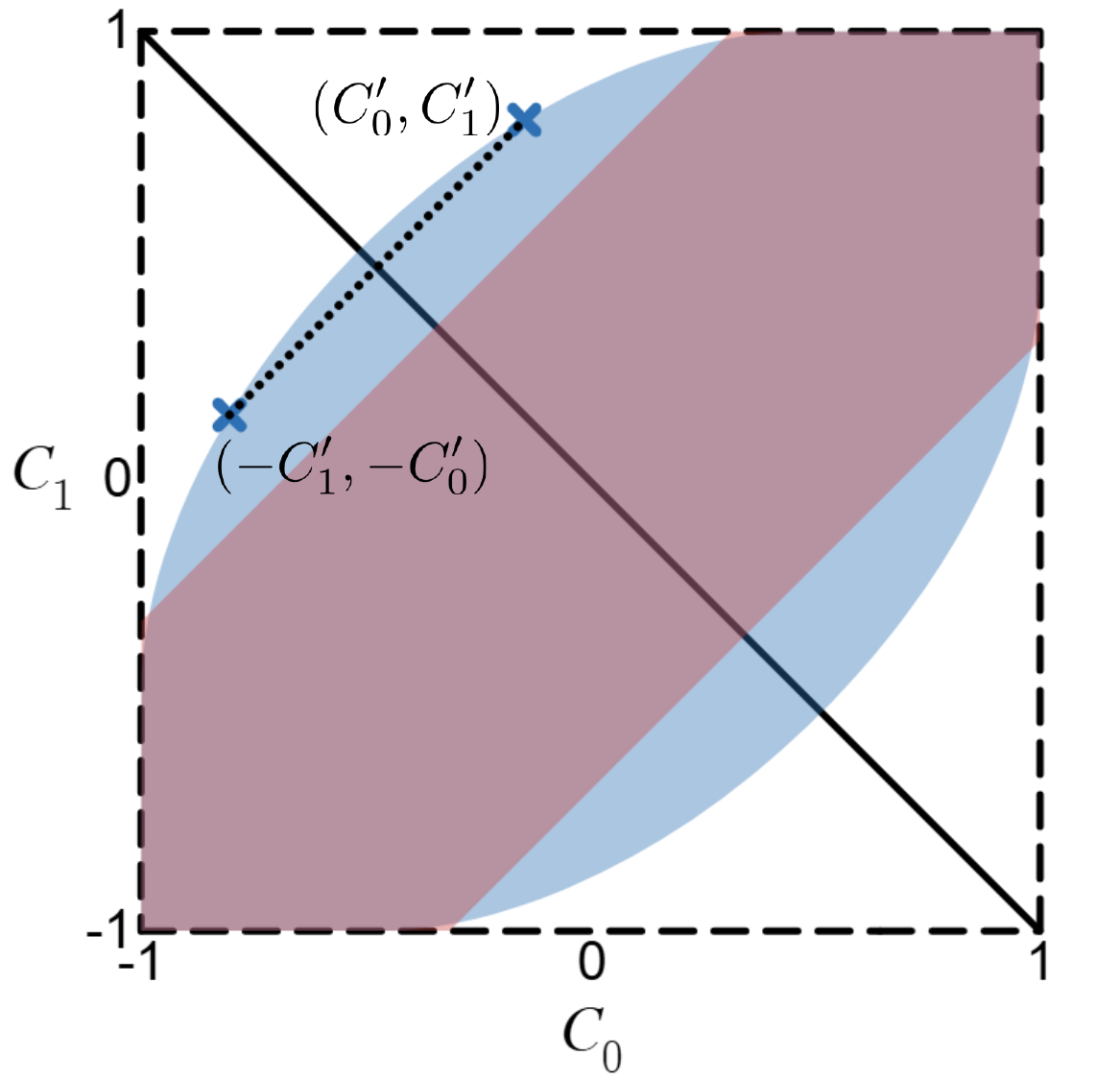}
\caption{On the boundary of $\mathcal{Q}_{\E,\Delta t}$, we have the points $(C'_0,C'_1)$ and $(-C'_1,-C'_0)$ on the line $L_{C'_0-C'_1}$. The intersection of the dotted line (given by $L_{C'_0-C'_1}$) and the diagonal $C_1=-C_0$ is the point $C^\star$. The value $H^\star(\bm{C}^\star)$ lower-bounds $H^\star$ everywhere on this line.}
\label{fig:H_symmetry}
\end{figure}

For further results about this optimisation problem, let us explicitly denote the dependence of the number of certified random bits $H^\star$ on the correlation $\bm{C}=(C_0,C_1)$ and $\mathfrak{E}$ by
\[
   H_{\rm dual}^\star=H^\star=H^\star(C_0,C_1,\mathfrak{E})=H^\star(\bm{C},\mathfrak{E}).
\]
From the primal formulation~(\ref{opt_prob}) in the main text, we first infer the following:
\begin{lemma}
\label{LemHConvex}
$H^\star$ is convex in the correlation $\bm{C}$, i.e.
\[
   H^\star(q\bm{C}+(1-q)\mathbf{C}',\mathfrak{E})\leq q H^\star(\bm{C},\mathfrak{E})+(1-q)H^\star(\mathbf{C}',\mathfrak{E})
\]
for all $0\leq q\leq 1$ and all correlations $\bm{C},\mathbf{C}'\in\mathcal{Q}_{\mathfrak{E},1}$.
\end{lemma}
\begin{proof}
Consider the optimal ensemble $\{p(\lambda),\bm{C}^\lambda,\mathfrak{E}^\lambda\}$ for which $H^\star(\bm{C},\mathfrak{E})$ is attained in~(\ref{opt_prob}). Without loss of generality, suppose that the optimal ensemble for $\mathbf{C}'$ is attained on the same set of hidden variables $\Lambda\ni\lambda$ (otherwise, use the union of both sets, and declare all additional elements to have zero probability), such that it can be denoted $\{\bar p(\lambda),\mathbf{\bar C}^\lambda,\mathfrak{\bar E}^\lambda\}$. For $i\in\{1,2\}$ and $\lambda\in\Lambda$, define $p(i,\lambda):=q p(\lambda)$ if $i=1$ and $(1-q)\bar p(\lambda)$ if $i=2$. Set $\bm{C}^{1,\lambda}:=\bm{C}^\lambda$ and $\bm{C}^{2,\lambda}:=\mathbf{\bar C}^\lambda$, as well as $\mathfrak{E}^{1,\lambda}:=\mathfrak{E}^\lambda$ and $\mathfrak{E}^{2,\lambda}:=\mathfrak{\bar E}^{\lambda}$. Then the ensemble $\{p(i,\lambda),\bm{C}^{i,\lambda},\mathfrak{E}^{i,\lambda}\}_{i,\lambda}$ satisfies
\begin{eqnarray*}
\sum_{i,\lambda}p(i,\lambda)\bm{C}^{i,\lambda}&=&\sum_\lambda q p(\lambda)\bm{C}^\lambda+\sum_\lambda (1-q)\bar p(\lambda)\mathbf{\bar C}^\lambda \\&=& q\bm{C}+(1-q)\mathbf{C}',\\
\sum_{i,\lambda}p(i,\lambda)\mathfrak{E}^{i,\lambda}&=&\sum_\lambda q p(\lambda)\mathfrak{E}^\lambda+\sum_\lambda (1-q)\bar p(\lambda)\mathfrak{\bar E}^\lambda\\
&=& q\mathfrak{E}+(1-q)\mathfrak{E}=\mathfrak{E},\\
\bm{C}^{i,\lambda}&\in& \mathcal{Q}_{\mathfrak{E}^{i,\lambda},1}.
\end{eqnarray*}
Therefore, we obtain
\begin{eqnarray*}
H^\star(q\bm{C}+(1-q)\mathbf{C}',\mathfrak{E})&\leq&\sum_{i,\lambda}p(i,\lambda)H(\bm{C}^{i,\lambda})\\
&=& q H^\star(\bm{C},\mathfrak{E})+(1-q)H^\star(\mathbf{C}',\mathfrak{E}).
\end{eqnarray*}
The claim follows.
\end{proof}
The following symmetry property is evident in the plot of Fig.~\ref{fig:min-entropy}:
\begin{lemma}{\label{lemmaSymmetryCorr}}
For all $0\leq\mathfrak{E}\leq \frac\pi 2$ and all $\bm{C}\in\mathcal{Q}_{\mathfrak{E},1}$, we have
\[
   H^\star(C_0,C_1,\mathfrak{E})=H^\star(C_1,C_0,\mathfrak{E})=H^\star(-C_0,-C_1,\mathfrak{E}).
\]
\end{lemma}
\begin{proof}
By definition of $H$, we have $H(C_0,C_1)=H(-C_0,-C_1)=H(C_1,C_0)$. Moreover, from the characterisation~(\ref{quantum-pure}), we know that $\bm{C}\in\mathcal{Q}_{\mathfrak{E},\Delta t}$ if and only if $-\bm{C}\in\mathcal{Q}_{\mathfrak{E},\Delta t}$ if and only if $(C_1,C_0)^\top\in\mathcal{Q}_{\mathfrak{E},\Delta t}$. Therefore, replacing $\bm{C}$ by $-\bm{C}$ and all $\bm{C}^\lambda$ by $-\bm{C}^\lambda$ in the primal problem~(\ref{opt_prob}) leaves the value of $H^\star$ invariant. And so does replacing $\bm{C}$ by $(C_1,C_0)^\top$ and all $\bm{C}^\lambda$ by $(C_1^\lambda,C_0^\lambda)^\top$.
\end{proof}
Note that the lemma also implies that
\[
   H^\star(C_0,C_1,\mathfrak{E})=H^\star(-C_1,-C_0,\mathfrak{E})
\]
which will be useful in the following lemma.
\begin{lemma}\label{LemMidpoint}
We have the inequality
\begin{eqnarray*}
    H^\star\left(\frac{C_0-C_1}2,-\frac{C_0-C_1}2,\mathfrak{E}\right)&\leq& H^\star(C_0,C_1,\mathfrak{E})\\
    &=& H^\star (-C_1,-C_0,\mathfrak{E}).
\end{eqnarray*}
\end{lemma}
\begin{proof}
Note that
\[
   \left(\begin{array}{c} \frac{C_0-C_1}2 \\ -\frac{C_0-C_1} 2 \end{array}\right)=\frac 1 2\left(\begin{array}{c}C_0 \\ C_1 \end{array}\right)+\frac 1 2 \left(\begin{array}{c} -C_1 \\ -C_0\end{array}\right)
\]
and use the previous two lemmas.
\end{proof}
This lemma tells us the following. In the correlation square (or rather in its subset $\mathcal{Q}_{\mathfrak{E},1}$ for a fixed value of $\mathfrak{E}$), consider the lines
\[
   L_\delta:=\{\bm{C}\in\mathcal{Q}_{\mathfrak{E},1}\,\,|\,\, C_0-C_1=\delta\}.
\]
Then $C^\star:=\left(\begin{array}{c} \frac{C_0-C_1}2 \\ -\frac{C_0-C_1} 2 \end{array}\right)$ is the center of the line $L_{C_0-C_1}$, and the value of $H^\star$ at this midpoint lower-bounds the value of $H^\star$ everywhere else on the line. In particular, if we have computed a certified lower bound to $H^\star(C^\star,\mathfrak{E})$, then this is a certified lower bound for all other correlations on this line, as illustrated by the dotted line of Fig.~\ref{fig:H_symmetry}. This is consistent with the plot in Fig.~\ref{fig:min-entropy}.

Indeed, we conjecture -- and the numerics suggest -- that $H^\star$ is constant on those lines, i.e.\ that we have equality in Lemma~\ref{LemMidpoint}. However, we have not managed to prove this.

\begin{lemma}{\label{lemma_tvec}}
For some $\bm{C}$ with $|C_0|\neq 1$ and $|C_1|\neq 1$, and some $\mathfrak{E}<\frac\pi 2$, consider the optimisation problem
\[
H_{\rm dual}^\star:=\sup_{\bm{t}\in \mathbb{R}^3} t(\bm t)+\bm{t}\cdot\left(\begin{array}{c} \bm{C} \\ \mathfrak{E}\end{array}\right),
\]
where
\begin{equation}
   t(\bm t):=\min_{(\bm{C}',\mathfrak{E}')\in\mathcal{Q}} \left[\strut H(\bm{C}')-\bm{t}\cdot\left(\begin{array}{c}\bm{C}'\\\mathfrak{E}'\end{array}\right)\right].
   \label{eqMinimizationInside}
\end{equation}
Then the $\bm{t}$ at which the maximum is attained has the properties that $t_1$ and $t_2$ have opposite signs, and that $t_3\leq 0$.

Furthermore, consider the special case $C_0\geq 0$ and $|C_1|\leq C_0$. Then there is always a maximiser $\bm{t}$ with $t_1\geq 0$, $t_2\in [-t_1,0]$ and
\[
   |t_3|\leq \min\left\{ \frac{h(C_1)+(C_0-C_1)t_1}{\mathfrak{E}},\frac{h(C_0)+(C_0-C_1)|t_2|}{\mathfrak{E}}\right\}
\]
(while still $t_3\leq 0$). Moreover, for every $\bm{C}'\in\mathcal{Q}_{\mathfrak{E},1}$ with $C'_0>C_0$ and $C'_1<C_1$, the range of optimisation is further restricted by
\[
   |t_1|(C'_0-C_0)+|t_2|(C_1-C'_1)\leq H(\bm{C}').
\]
\end{lemma}
\begin{proof}
Define
\[g_{\bm t}(\mathfrak{E}'):=\min_{\bm{C}'\in\mathcal{Q}_{\mathfrak{E}',1}}\left(H(\bm{C}')-\left(\begin{array}{c} t_1 \\ t_2 \end{array}\right)\cdot\bm{C}'-t_3 \mathfrak{E}'\right).
\]
Suppose that $\mathfrak{E}'\leq\mathfrak{E}''$, then $\mathcal{Q}_{\mathfrak{E}',1}\subseteq \mathcal{Q}_{\mathfrak{E}'',1}$, and so
\begin{eqnarray*}
g_{\bm t}(\mathfrak{E}')&\geq&\min_{\bm{C}'\in\mathcal{Q}_{\mathfrak{E}'',1}}\left(H(\bm{C}')-\left(\begin{array}{c} t_1 \\ t_2 \end{array}\right)\cdot\bm{C}'-t_3 \mathfrak{E}'\right)\\
&=& g_{\bm{t}}(\mathfrak{E}'')+t_3(\mathfrak{E}''-\mathfrak{E}').
\end{eqnarray*}
Hence, if $t_3\geq 0$, we have $g_{\bm{t}}(\mathfrak{E}')\geq g_{\bm{t}}(\mathfrak{E}'')$, and so
\begin{eqnarray*}
t(\bm{t})&=&\min_{\mathfrak{E}'} g_{\bm{t}}(\mathfrak{E}')=g_{\bm{t}}(\frac\pi 2)\\
&=&\min_{\bm{C}'\in[-1,1]^2} \left(H(\bm{C}')-\left(\begin{array}{c} t_1 \\ t_2 \end{array}\right)\cdot\bm{C}'\right)-t_3\cdot\frac\pi 2.
\end{eqnarray*}
Therefore, if $t_3\geq 0$, then
{\small\[
   t(\bm{t})+\bm{t}\cdot\left(\begin{array}{c} \bm{C} \\ \mathfrak{E}\end{array}\right)=\min_{\bm{C}'\in[-1,1]^2} \left(H(\bm{C}')-\left(\begin{array}{c} t_1 \\ t_2 \end{array}\right)\cdot\bm{C}'\right)+t_3\left(\mathfrak{E}-\frac\pi 2\right)
\]}
which is maximised for $t_3=0$. Hence, for
the maximising $\bm{t}$, we have $t_3\leq 0$, and also
\begin{equation}
   H_{\rm dual}^\star\leq H(\mathbf{C}')-\mathbf{t}\cdot\left(\begin{array}{c}\mathbf{C}'\\ \mathfrak{E}'\end{array}\right)+\mathbf{t}\cdot\left(\begin{array}{c}\mathbf{C}\\ \mathfrak{E}\end{array}\right)
   \label{eqUseful}
\end{equation}
for all $(\bm{C}',\mathfrak{E}')\in\mathcal{Q}$. In particular, for $\mathfrak{E}'=0$ and $\bm{C}'=(1,1)$, we have
\[
   H_{\rm dual}^\star\leq t_1(C_0-1)+t_2(C_1-1)-|t_3|\mathfrak{E}.
\]
But $H_{\rm dual}^\star\geq 0$, and so $t_1$ and $t_2$ cannot both be positive. Similarly, choosing $\mathfrak{E}'=0$ and $\bm{C}'=(-1,-1)$ shows that they cannot both be negative.

Now let us assume that $C_0\geq 0$ and $|C_1|\leq C_0$. Due to the symmetry property
\[
   H\left(\begin{array}{c} C_0 \\ C_1 \end{array}\right)=H\left(\begin{array}{c} -C_0 \\ -C_1 \end{array}\right)=H\left(\begin{array}{c} C_1 \\ C_0 \end{array}\right)=H\left(\begin{array}{c} -C_1 \\ -C_0 \end{array}\right)
\]
the value of $t(\bm{t})$ is constant on all $\bm{t}\in T$, where
\[
   T=\left\{
      \left(\begin{array}{c} t_1\\t_2\\t_3\end{array}\right),\left(\begin{array}{c} -t_1\\-t_2\\t_3\end{array}\right),\left(\begin{array}{c} t_2\\t_1\\t_3\end{array}\right),\left(\begin{array}{c} -t_2\\-t_1\\t_3\end{array}\right)
   \right\}.
\]
Now consider the maximisation in the first line of our optimisation problem, renaming $\bm{t}$ to $\bm{t}'$ in order not to confuse it with our currently fixed vector $\bm{t}$. The expression $t(\bm{t}')$ is constant on all $\bm{t}'\in T$, and hence, the maximum over those four different $\bm{t}'$ is attained whenever $t'_1 C_0+t'_2 C_1$ is maximised. One of the four vectors in $T$ has the property that the first entry is non-negative and the second entry is in absolute value less than the first. Let us rename the vectors such that this is indeed $\bm{t}=(t_1,t_2,t_3)^\top$. Then, for example,
\begin{eqnarray*}
   \left(\begin{array}{c}t_1 \\ t_2\end{array}\right)\bm{C}-\left(\begin{array}{c} -t_1 \\ -t_2 \end{array}\right)\bm{C}&=&2t_1 C_0+2t_2 C_1\\
   &\geq& 2 t_1 C_0-2|t_2|\,|C_1|)\geq 0.
\end{eqnarray*}
Similar comparison with the other two entries shows that $\max_{\bm{t}'\in T}\left(\begin{array}{c} t_1 \\ t_2 \end{array}\right)\cdot\bm{C}=\left(\begin{array}{c} t_1 \\ t_2 \end{array}\right)\cdot\bm{C}$. Hence, the maximum in the first line of the optimisation problem will always be attained at some $\bm{t}$ with $t_1\geq 0$ and $|t_2|\leq t_1$. Since we have already shown that $t_2\leq 0$, it follows that $t_2\in [-t_1,0]$.

Now recall Eq.~(\ref{eqUseful}). For $\mathfrak{E}'=0$. Since $\mathcal{Q}_{0,\Delta t}=\{(a,a)^\top\,\,|\,\, -1\leq a \leq 1\}$, we obtain
\[
   0\leq H_{\rm dual}^\star\leq h(a)-t_1(a-C_0)-t_2(a-C_1)-|t_3|\mathfrak{E}'
\]
for all $a\in[-1,1]$, where $h(a)=-\frac{1-a}2 \log_2\frac{1-a}2-\frac{1+a}2\log_2\frac{1+a}2$. The special cases $a=C_0$ and $a=C_1$ prove the bound on $|t_3|$ in the statement of then lemma. The final bound follows from the special case $\mathfrak{E}'=\mathfrak{E}$ in~(\ref{eqUseful}).
\end{proof}

\subsection{Discretisation and error bounds}
\label{SubsecDiscretisation}

While the primal optimisation problem~(\ref{eq:primal}) has an in principle unbounded number of free variables (since there is an arbitrary number of $\lambda$), the dual problem~(\ref{eq:dual}) optimises over a finite and small number of parameters. It hence admits an approximation by discretisation in a meaningful way, which is where we are heading next.

Restricting the domain of a maximisation can only decrease the optimal value and, therefore, preserves the lower-bound property. This will be the case in Lemma~\ref{LemFirstDiscr} below. However, discretising a minimisation may lead to a larger value if the true minimum is missed, which is a problem that we deal with in Lemmas~\ref{LemDiscrE} and~\ref{lemmaCertified_lb} below, when approximating $t(\bm{t})$ from below.

To mediate this problem, we will determine a strict bound as to by how much the actual value of $t(\bm{t})$ is overestimated, and subtract it from the naive numerical estimate. The subtracted correction will vanish in the limit of better and better discretisation.

To bound the overestimation, we essentially use variants of what is known as an application of the \textit{mean value theorem} in calculus. In the simplest case, suppose we have a real function $f:[a,b]\to\mathbb{R}$, where $[a,b]\subset\mathbb{R}$ is some interval, and we know that $|f'(x)|\leq c$ for all $x\in [a,b]$. Then it follows that $|f(x_1)-f(x_0)|\leq c|x_1-x_0|$. Searching the minimum of $f(x)$ on a discrete grid of distance $2\varepsilon$ between points will thus overestimate the true minimum by at most $c\varepsilon$. Here we use variations of this idea in more than one dimension, and when $c$ is not necessarily determined by computing a derivative.

Let us first discretise the choice of vectors $\bm{t}$. Rather than from choosing them from all of $\mathbb{R}^3$, let us restrict our attention to the cube $[-L,L]^3$, where $L\in\mathbb{N}$ is some (large) integer. Moreover, let us discretise the interval $[L,L]$ into discrete points
{\small\[
   I_{L,M}:=\left\{ -\frac{ML}M,\frac{-ML+1}M,\frac{-ML+2} M,\ldots,0,\frac 1 M,\ldots,\frac{ML}M\right\}
\]}
where $M\in\mathbb{N}$ is also some (large) integer. This defines a new optimisation problem whose solution lower-bounds the dual problem:
\begin{lemma}
\label{LemFirstDiscr}
Consider the optimisation problem
\[
H_{\rm dual}^\star(L,M):=\max_{\bm{t}\in I_{L,M}^3} t(\bm t)+\bm{t}\cdot\left(\begin{array}{c} \bm{C} \\ \mathfrak{E}\end{array}\right),
\]
where
\begin{equation}
   t(\bm t):=\min_{(\bm{C}',\mathfrak{E}')\in\mathcal{Q}} \left[\strut H(\bm{C}')-\bm{t}\cdot\left(\begin{array}{c}\bm{C}'\\\mathfrak{E}'\end{array}\right)\right].
   \label{eqMintt}
\end{equation}
Then $H_{\rm dual}^\star(L,M)\leq H_{\rm dual}^\star$ for all $L,M$, and $\lim_{L\to\infty}\lim_{M\to\infty} H_{\rm dual}^\star(L,M)=H_{\rm dual}^\star=H^\star$.
\end{lemma}
\begin{proof}
Increasing the domain of maximisation from $\bm{t }\in I_{L,M}^3$ to $\mathbb{R}^3$ reproduces the optimisation problem~(\ref{eq:dual}), and hence, $H_{\rm dual}^\star$ upper-bounds $H_{\rm dual}^\star(L,M)$. Also, by continuity of the function to be optimised, it is clear that in the limit of better and better discretisation of the cube, and then of extending the cube to infinity, we reproduce $H^\star_{\rm dual}$.
\end{proof}
Now we discretise the problem in the energy uncertainty $\mathfrak{E}'$. For integers $L,N\in\mathbb{N}$, define
\[
   E_{L,N}:=\left\{\frac 1 {LN},\frac 2 {LN},\ldots,\frac{LN-1}{LN},1\right\}\cdot\frac\pi 2 \subset [0,\frac\pi 2].
\]
Then we have:
\begin{lemma}\label{LemDiscrE}
For every $L,M,N\in\mathbb{N}$, the solution of
\begin{eqnarray*}
H_{\rm dual}^\star(L,M,N):=\max_{\bm{t}\in I_{L,M}^3} t_{L,N}(\bm t)-|t_3|\frac{\pi/2}{LN}+\bm{t}\cdot\left(\begin{array}{c} \bm{C} \\ \mathfrak{E}\end{array}\right),\\
t_{L,N}(\bm t):=\min_{\mathfrak{E}'\in E_{L,N}} \min_{\bm{C}'\in\mathcal{Q}_{\mathfrak{E}',\Delta t=1}} \left[\strut H(\bm{C}')-\bm{t}\cdot\left(\begin{array}{c} \bm{C}'\\ \mathfrak{E}'\end{array}\right)\right]
\end{eqnarray*}
satisfies
\[
   H_{\rm dual}^\star(L,M,N)\leq H_{\rm dual}^\star(L,M)\leq H^\star,
\]
and so the left-hand side gives a certifiable lower bound to $H^\star$. Moreover,
\[
   \lim_{N\to\infty}\lim_{L\to\infty}\lim_{M\to\infty} H_{\rm dual}^\star(L,M,N)=H^\star.
\]
\end{lemma}
\begin{proof}
Pick any $\bm{t}\in I_{L,M}^3$, and let $(\mathbf{C}',\mathfrak{E}')\in\mathcal{Q}$ be any minimiser in~(\ref{eqMintt}). Let $\mathfrak{\tilde E}'$ be the smallest number in the set $E_{L,N}$ which is not smaller than $\mathfrak{E}'$, then $\mathfrak{E}'\leq \mathfrak{\tilde E}'$ and $\mathfrak{\tilde E}'-\mathfrak{E}'\leq\frac{\pi/2}{LN}$. Moreover, $\mathcal{Q}_{\mathfrak{E}',1}\subseteq \mathcal{Q}_{\mathfrak{\tilde E}',1}$, and so
\begin{eqnarray*}
   t_{L,N}(\bm{t})&\leq& H(\bm{C}')-\bm{t}\cdot\left(\begin{array}{c}\bm{C}'\\ \mathfrak{\tilde E}'\end{array}\right)=t(\bm{t})+\bm{t}\cdot\left(\begin{array}{c} 0 \\ \mathfrak{E}'-\mathfrak{\tilde E}'\end{array}\right)\\
   &=& t(\bm{t})+t_3(\mathfrak{E}'-\mathfrak{\tilde E}')\leq t(\bm{t})+|t_3|\cdot\frac{\pi/2}{LN}.
\end{eqnarray*}
Thus, $H_{\rm dual}^\star(L,M,N)\leq H_{\rm dual}^\star(L,M)$, and all further claims follow from the previous lemmas.
\end{proof}
Finally, we provide a discretisation for the optimisation over the set $\mathcal{Q}_{\mathfrak{E}',1}$. One crucial insight below will be that it is sufficient to optimise over the correlations $\bm{C}'$ that lie on the boundary curves of $\mathcal{Q}_{\mathfrak{E}',\Delta t=1}$. We will discretise these curves by introducing another integer $S\in\mathbb{N}$ and setting
\begin{eqnarray*}
C_2^{(\mathfrak{E}')}=\left\{\frac\Delta{S},\frac {2\Delta}{S},\ldots,\frac{S\Delta}{S}\right\},\quad
C_1^{(\mathfrak{E}')}=C_2^{(\mathfrak{E}')}+\mathfrak{E}',
\end{eqnarray*}
where $\Delta=\frac\pi 2 - \mathfrak{E}'$. The two curves that bound the set $\mathcal{Q}_{\mathfrak{E}',1}$ are
{\small\begin{eqnarray*}
c_1^{(\mathfrak{E}')}(s)&=&(2\cos^2 s-1,2\cos^2(s-\mathfrak{E}')-1)  \quad (\mathfrak{E}'\leq s \leq\frac\pi 2),\\
c_2^{(\mathfrak{E}')}(s)&=& (2 \cos^2 s -1,2 \cos^2(s+\mathfrak{E}')-1)\quad (0\leq s \leq \frac\pi 2 -\mathfrak{E}'),
\end{eqnarray*}}
which defines some of the notation used in the following lemma.
\begin{lemma}{\label{lemmaCertified_lb}}
For every $L,M,N,S\in\mathbb{N}$, the solution of
{\small \begin{align*} H_{\rm dual}^\star(L,M,N,S)&:=\max_{\bm{t}\in I_{L,M}^3} t_{L,N,S}(\bm t)-|t_3|\frac{\pi/2}{LN}+\bm{t}\cdot\left(\begin{array}{c} \bm{C} \\ \mathfrak{E}\end{array}\right),\\ t_{L,N,S}(\bm t)
&:=\min_{\substack{\mathfrak{E}'\in E_{L,N} \\ \bm{C}'\in \mathcal{Q}_{\mathfrak{E}',1}}}\Bigl[ \strut \{-t_1-t_2-t_3\mathfrak{E}',t_1+t_2-t_3\mathfrak{E}'\}\Bigr.\\ 
&\cup\{H(c_1^{(\mathfrak{E}')}(s))-\bm{t}\cdot\left(\begin{array}{c} c_1^{(\mathfrak{E}')}(s)\\ \mathfrak{E}'\end{array}\right)-\delta_{\bm t}\,\,|\,\, s\in C_1^{(\mathfrak{E}')}\}\\ 
&\Bigl. \cup\{H(c_2^{(\mathfrak{E}')}(s))-\bm{t}\cdot\left(\begin{array}{c} c_2^{(\mathfrak{E}')}(s)\\ \mathfrak{E}'\end{array}\right)-\delta_{\bm t}\,\,|\,\, s\in C_2^{(\mathfrak{E}')}\}\Bigr],
\end{align*}}
where $\delta_{\bm{t}}:=\frac{(\frac\pi 2-\mathfrak{E}')(1+2|t_1|+2|t_2|)}{S}$,
satisfies
\[
   H_{\rm dual}^\star(L,M,N,S)\leq H_{\rm dual}^\star(L,M,N),
\]
and $\lim_{S\to\infty} H_{\rm dual}^\star(L,M,N,S)=H_{\rm dual}^\star(L,M,N)$. Hence
\[
   H_{\rm dual}^\star(L,M,N,S)\leq H^\star,
\]
and
\[
   \lim_{S\to\infty} \lim_{N\to\infty}\lim_{L\to\infty}\lim_{M\to\infty}H_{\rm dual}^\star(L,M,N,S)=H^\star.
\]
\end{lemma}
\begin{proof}
The function $f(\bm{C}'):=H(\bm{C}')-\bm{t}\cdot\left(\begin{array}{c} \bm{C}'\\ \mathfrak{E}'\end{array}\right)$ is concave and continuous in $\bm{C}'$. Hence, its minimum on the compact convex set $\mathcal{Q}_{\mathfrak{E}',1}$ is attained on the extreme points of this set. These are
\begin{eqnarray*}
   \partial_{\rm ext}\mathcal{Q}_{\mathfrak{E}',1}&=&\left\{\left(\begin{array}{c}-1 \\ -1\end{array}\right),\left(\begin{array}{c}1 \\ 1\end{array}\right)\right\}\cup \left\{c_2^{(\mathfrak{E}')}(s)\,\,|\,\,0\leq s \leq \frac \pi 2\right\}\\
   && \cup \left\{c_1^{(\mathfrak{E}')}(s)\,\,|\,\, E\leq s \leq \frac \pi 2\right\}.
\end{eqnarray*}
It can be checked that
\begin{eqnarray*}
   \left|\frac\partial {\partial s}H(c_1^{(\mathfrak{E})}(s))\right|\leq 2,\\
   \left|\frac\partial{\partial s} \bm{t}\cdot\left(\begin{array}{c} c_1^{(\mathfrak{E})}(s) \\ \mathfrak{E}'\end{array}\right)\right|\leq 4(|t_1|+|t_2|).
\end{eqnarray*}
To see the first inequality, some tedious but straightforward calculus shows that $|(\ln(\tan^2(x))) \sin(2x)|\leq 1.33<2\ln 2$ for all $x\in\mathbb{R}$. Furthermore
{\small \[
   \frac\partial{\partial s}H(c_1^{(\mathfrak{E})}(s))=\frac{\left(\ln\tan^2(\mathfrak{E}-s)\right)\sin(2(\mathfrak{E}-s))+\left(\ln\tan^2 s\right)\sin(2s)}{2\ln 2}
\]}
which proves the first inequality on the derivative; the second one is straightforward.
For all $s\in[\mathfrak{E},\frac\pi 2]$ there exists some $s'\in C_1^{(\mathfrak{E})}$ with $|s-s'|\leq \frac \Delta {2S}$. Since $|\frac\partial{\partial s}f(c_1^{(\mathfrak{E})}(s))|\leq 2+4(|t_1|+|t_2|)$, we have $|f(c_1^{(\mathfrak{E})}(s'))-f(c_1^{(\mathfrak{E})}(s))|\leq\frac{\Delta(1+2|t_1|+2|t_2|)}{S}=\delta_{\bm{t}}$. Thus
\[
   \min_{s\in C_1^{(\mathfrak{E}')}}f(c_1^{(\mathfrak{E}')}(s))-\delta_{\bm{t}}\leq \min_{s\in[\mathfrak{E}',\frac\pi 2]}f(c_1^{(\mathfrak{E}')}(s)),
\]
and similar argumentation applies to the other curve. Hence $t_{L,N,S}(\bm{t})\leq t_{L,N}(\bm{t})$, and the claims follows.
\end{proof}

\subsection{Numerical algorithm implementation of the discretised dual bound optimisation}
{\label{sec-algorithm}}

In this subsection, we describe how our optimisation procedure established in Lemma~\ref{lemmaCertified_lb} can be concretely implemented on a computer. To obtain the plot in Figure~\ref{fig:min-entropy}, we have done this together with one further optimisation that we describe in Subsection~\ref{SubsecDiagonal} below.

Given some observed point $\bm{x}_0 = (C_0,C_1,\mathfrak{E})$, our goal is to evaluate the lower bound
\begin{equation}
   H^\star_{\mathrm{dual}}(L,M,N,S) :=
   \max_{\bm{t}=(t_1,t_2,t_3)}
   \left[
      t_{L,N,S}(\bm{t}) - |t_3|\,\tfrac{\pi/2}{LN} + \bm{t}\cdot\bm{x}_0
   \right],
\end{equation}
where $t_{L,N,S}(\bm{t})$ is the discretised version of $t(\bm{t})$ as defined in Lemma~\ref{lemmaCertified_lb}. Lemma~\ref{lemmaCertified_lb} also tells us that, for every choice $L,M,N,S\in\mathbb{N}$, one has
\[
   H^\star_{\mathrm{dual}}(L,M,N,S)\leq H^\star,
\]
so any value obtained in this way gives a certified lower bound on the true entropy. Increasing $L,M,N,S$ increases the tightness and in the asymptotic limit $H^\star_{\mathrm{dual}}(L,M,N,S)$ converges to $H^\star$, at the cost of higher runtime.

In practice, we use symmetry properties and reduction lemmas to reduce the runtime by restricting both the input point and the dual variables. The algorithm proceeds as follows:
\begin{enumerate}
    \item \textbf{Canonical correlation sector. }Using the symmetries of Lemma~\ref{LemMidpoint} and Lemma~\ref{lemmaSymmetryCorr}, we first map the observed correlations $(C_0,C_1)$ into the canonical sector
\[
   C_0\geq 0,\qquad |C_1|\leq C_0.
\]
All symmetry-related points have the same value of $H^\star_{\mathrm{dual}}(L,M,N,S)$, so restricting to this sector does not change the optimisation problem.
\item \textbf{Restricted discretised $t$–grid. }We discretise the cube $[-L,L]^3$ using the 1D grid
\[
   I_{L,M}
   :=\left\{-L, -L+\frac{1}{M}, \dots, L-\frac{1}{M}, L\right\},
\]
and consider $\bm{t}$ only on the Cartesian product $I_{L,M}^3$. Then, using Lemma~\ref{lemma_tvec} we further restrict the search to scan over those $\bm{t}\in I_{L,M}^3$ fulfilling
\[
   t_3\leq 0,\qquad t_1\geq 0,\qquad |t_2|\leq t_1.
\]

\item \textbf{Discretisation of the energy uncertainty.} For each fixed $\bm{t}$, we need to evaluate the inner minimum over the energy uncertainty $\mathfrak{E}'$ and the quantum set $\mathcal{Q}_{\mathfrak{E}',1}$:
\[
   t_{L,N}(\bm{t}) :=
   \min_{\mathfrak{E}'\in E_{L,N}}
   \min_{\bm{C}'\in\mathcal{Q}_{\mathfrak{E}',1}}
   \bigl[H(\bm{C}')-\bm{t}\cdot(\bm{C}',\mathfrak{E}')\bigr],
\]
where
\[
   E_{L,N} :=
   \left\{
      \frac{1}{LN},\frac{2}{LN},\dots,\frac{LN-1}{LN},1
   \right\}\frac{\pi}{2}
   \subset(0,\tfrac{\pi}{2}].
\]
In practice, this is implemented by scanning over the grid $\mathfrak{E}' \in E_{L,N}$.

The discretisation error from replacing the continuous interval $[0,\tfrac{\pi}{2}]$ by $E_{L,N}$ is compensated by the additive penalty $ -|t_3|\frac{\pi/2}{LN}$,
which is the term explicitly subtracted in the definition of $H^\star_{\mathrm{dual}}(L,M,N,S)$. 

\item \textbf{Discretisation of the quantum boundary and $S$–dependent error.} For each $\mathfrak{E}'\in E_{L,N}$ and dual vector $\bm{t}$, the relevant minimum over $\bm{C}'$ is attained on the extremal boundary of $\mathcal{Q}_{\mathfrak{E}',1}$, consisting of:
\begin{itemize}
  \item the two corners $(-1,-1)$ and $(1,1)$, and
  \item the two boundary curves
  \[
     c_1^{(\mathfrak{E}')}(s),\qquad
     c_2^{(\mathfrak{E}')}(s),
  \]
  where the parametrisations $c_1^{(\mathfrak{E}')}(s)$, $c_2^{(\mathfrak{E}')}(s)$ correspond to the probability-space parametrisations $p_1(\tau)$, $p_2(\tau)$ of Section~\ref{quantum-model} through the change of variables $s=\mathfrak{E}'\tau$ and using $C_x = 2P_x^+ - 1$.
\end{itemize}
The corners are evaluated exactly, via
\[
   f_{\mathrm{corner},1} = -\bm{t}\cdot((-1,-1),\mathfrak{E}'),
   \qquad
   f_{\mathrm{corner},2}  = -\bm{t}\cdot((1,1),\mathfrak{E}'),
\]
since $H(\pm 1,\pm 1)=0$.

The contribution from the curves is approximated by sampling $s$ on a uniform grid of size $S$,
\[
   s\in C_1^{(\mathfrak{E}')},\qquad s\in C_2^{(\mathfrak{E}')},
\]
where if $s\in [\mathfrak{E}',\pi/2]$ the step size is $s_k = \mathfrak{E}'+k\frac{\Delta}{S} $ for the first curve, with $\Delta := \frac{\pi}{2}-\mathfrak{E}'$ and $ k=1,\dots,S$, and if $s\in[0,\pi/2-\mathfrak{E}']$ then the step size is $\ s_k = k\frac{\Delta}{S}$.

For each sample we evaluate
\[
   f(c_i^{(\mathfrak{E}')}(s_k),\mathfrak{E}') := H(c_i^{(\mathfrak{E}')}(s_k))
     - \bm{t}\cdot(c_i^{(\mathfrak{E}')}(s_k),\mathfrak{E}'),
\]
and then subtract a $t$–dependent error term
\[
   \delta_{\bm{t}}(\mathfrak{E}')
   :=
   \frac{\bigl(\tfrac{\pi}{2}-\mathfrak{E}'\bigr)\bigl(1+2|t_1|+2|t_2|\bigr)}{S},
\]
which upper-bounds the difference between the discrete minimum over the $S$ grid points and the true minimum along the continuous curve.

Altogether, for fixed $\bm{t}$ and $\mathfrak{E}'$, we compute
\begin{eqnarray}
   f_{\min}(\bm{t},\mathfrak{E}') :=
   \min&\Bigl\{&
      f_{\mathrm{corner},1},\,
       f_{\mathrm{corner},2},\, \nonumber\\
      && \min_k f(c_1^{(\mathfrak{E}')}(s_k),\mathfrak{E}') - \delta_{\bm{t}}(\mathfrak{E}'), \nonumber\\
      && \min_k f(c_2^{(\mathfrak{E}')}(s_k),\mathfrak{E}') - \delta_{\bm{t}}(\mathfrak{E}') \nonumber
   \Bigr\},
\end{eqnarray}
to finally obtain
\[
   t_{L,N,S}(\bm{t}) :=
   \min_{\mathfrak{E}'\in E_{L,N}} f_{\min}(\bm{t},\mathfrak{E}').
\]

\item \textbf{Dual value at the observed point and global maximisation.} For each scanned $\bm{t}$, we combine all discretisation corrections and evaluate the dual value at the observed point. That is,
\[
   D_{L,N,S}(\bm{t};\bm{x}_0)
   :=
   t_{L,N,S}(\bm{t}) - |t_3|\,\tfrac{\pi/2}{LN}
   + \bm{t}\cdot\bm{x}_0.
\]
The algorithm iterates over all $\bm{t}$ in the restricted search grid $\tilde{I}_{L,M}^3 $ and keeps track of the maximum value found
\[
   H^\star_{\mathrm{dual}}(L,M,N,S)
   = \max_{\bm{t}\in \tilde{I}_{L,M}^3 } D_{L,N,S}(\bm{t};\bm{x}_0).
\]

Finally, since $H^\star\geq 0$, we define the certified lower bound returned by the algorithm as
\[
   H_{\mathrm{cert}}(L,M,N,S) := 
   \max\bigl\{0,\ H^\star_{\mathrm{dual}}(L,M,N,S)\bigr\}.
\]
\end{enumerate}

This procedure is guaranteed to lower-bound $H^\star_{\textrm{dual}}$, thus enabling randomness certification under the semi-device-independent assumptions of the main text.

\subsubsection*{Choice of discretisation parameters.}
The natural numbers $L,M,N,S$ control different aspects of the approximation:
\begin{itemize}
  \item $L$ and $M$ determine the size and resolution of the dual grid $I_{L,M}^3$. They are the most expensive in runtime since the number of $\bm{t}$ points in the grid grow on the order of $L^3M^3$. We heuristically find $L=20$ to yield the tightest $H_{\mathrm{dual}}^\star$ values and we compromise to values $M\leq 5$ for runtime speed-up. For a quick test, lower values significantly decrease the runtime (as well as the tightness).
  \item $N$ controls the resolution of the energy grid $E_{L,N}$ and appears linearly in the cost (through the scan over $\mathfrak{E}'$). Heuristically, we find that values on the order of $10^3$ provide a good trade-off between tightness and runtime, and $N$ increasing above the order of $10^4$ yield small increments in tightness.
  \item $S$ controls the resolution of the boundary discretisation on each curve and also appears linearly in the cost. Heuristically, we tend to set $S=N$.
\end{itemize}

\subsection{Efficiently generating the numerical plot via the antisymmetric diagonal}
\label{SubsecDiagonal}

This subsection describes an observation which allows us to compute the plot in Figure~\ref{fig:min-entropy} in an even more efficient way. In the previous subsection, we have described how to evaluate the quantity
\[
   H_{\mathrm{cert}}(L,M,N,S)
   = \max\{0,\,H^\star_{\mathrm{dual}}(L,M,N,S)\}
\]
for a given observed point $\bm{x}_0 = (C_0,C_1,\mathfrak{E})$.  
To efficiently generate the full plot shown in Fig.~\ref{fig:min-entropy}, we exploit the symmetry and convexity lemmas to avoid scanning the entire $(C_0,C_1)$–plane.  

The key observation is Lemma~\ref{LemMidpoint}, which states that for every fixed $\mathfrak{E}$ the quantity $H^\star$ satisfies
\[
   H^\star\!\left(\frac{C_0-C_1}{2},-\frac{C_0-C_1}{2},\mathfrak{E}\right)
   \;\leq\;
   H^\star(C_0,C_1,\mathfrak{E}),
\]
and that the point
\[
   C^\star :=
   \left(\frac{C_0-C_1}{2},-\frac{C_0-C_1}{2}\right)
\]
lies on the diagonal $C_1=-C_0$.

Thus the value on the diagonal provides a certified lower bound for every point on the line
\[
   L_\delta := \{(C_0,C_1)\mid C_0-C_1=\delta\},
\]
since $C^\star$ is the midpoint of $L_\delta$.  
In other words, every line with constant $C_0-C_1$ is lower-bounded by its intersection with the antisymmetric diagonal given by $C_0=-C_1$. Indeed, due to our numerical results, we conjecture that equality holds in Lemma~\ref{LemMidpoint}, so that the value of $H^\star$ is constant on those lines and equal to that of the midpoint.

This allows us to generate the full 2D plot using only a 1D sweep along the diagonal to obtain $H^\star_{\rm dual}(C_0,-C_0,\mathfrak{E})$ and then propagate this value along the corresponding $L_\delta$-lines (for all $(C_0,C_1)$ inside the quantum set). This diagonal-only method greatly reduces the runtime while preserving the validity of the certified lower bounds.\\

\subsection{Maximal certified $H_{\mathrm{dual}}^\star$ found for $\mathfrak{E}\cdot\Delta t=1/2$.}
{\label{sec-maxH}}

For the special case $\mathfrak{E}\cdot\Delta t=1/2$ plotted in Fig.~\ref{fig:min-entropy} we additionally performed a higher-precision evaluation of the discretised dual along the antisymmetric diagonal $C_1 = -C_0$. Numerically, the most optimal value we could find appears at $(C_0,C_1)=(\pm \sin(\mathfrak{E}\Delta t),\mp \sin(\mathfrak{E}\Delta t))$. Indeed, Lemma~\ref{LemHConvex} implies that the maximal value of $H^\star$ on the diagonal $C_1=-C_0$ is found at these points.

For the point $(C_0,C_1)=(\sin(1/2),-\sin(1/2))$, running the algorithm of Sec.~\ref{sec-algorithm} to do a dual brute-force search with parameters $L=20$, $M=5$, $N=S=20000$, we obtained the dual vector
$\bm{t}_{\mathrm{best}}=(11.2,-11.2,-19.8)$, yielding the certified lower bound $H_{\mathrm{dual}}^\star(20,5,20000,20000) = 0.8113$.

While we believe this value is not yet fully tight, further increasing the grids improves the bound only marginally, indicating that this value is close to the true optimum.

\end{document}